\documentclass[a4paper,11pt]{article}
\pdfoutput=1
\usepackage{geometry}
\usepackage[svgnames]{xcolor}
\usepackage[colorlinks=true,urlcolor=black,linkcolor=blue,citecolor=blue,bookmarks=false]{hyperref}
\usepackage{amsfonts,amsmath,amssymb}
\usepackage{graphicx}
\usepackage{dsfont}
\usepackage{graphicx}
\usepackage{bm}% bold math
\usepackage[font=small,labelfont=bf]{caption}
\renewcommand*{\thepage}{\footnotesize\arabic{page}}
\usepackage{color}
\usepackage{enumitem}
\usepackage{boldline}
\usepackage{changepage}
\usepackage{cite}
\usepackage{textcomp}
\usepackage[title]{appendix}
\usepackage{url}
\usepackage[T1]{fontenc}
\usepackage{authblk}
\usepackage{breakurl}
\usepackage{fancyhdr}
\usepackage[export]{adjustbox}
\usepackage[percent]{overpic}
 \usepackage{mathpazo}
 \usepackage{multirow}
 \usepackage{float}
\usepackage{soul}
\setstcolor{red}

\title{\bf Collective Behavior in the North Rhine-Westphalia Motorway Network}

\author{Shanshan Wang \thanks{shanshan.wang@uni-due.de}, Sebastian Gartzke, Michael Schreckenberg and Thomas Guhr}
\affil{\textit{Faculty of Physics, University of Duisburg--Essen, Duisburg, Germany}}

\date{\today}
 
\begin{document}
\maketitle

\noindent {\bf Abstract.}  To understand the dynamics on complex networks, measurement of correlations is indispensable. In a motorway network, it is not sufficient to collect information on fluxes and velocities on all individual links, i.e.~parts of the freeways between ramps and highway crosses. The interdependencies and mutual connections are also of considerable interest.  We analyze correlations in the complete motorway network in North Rhine-Westphalia, the most populous state in Germany. We view the motorway network as a complex system consisting of road sections which interact via the motion of vehicles, implying structures in the corresponding correlation matrices. In particular, we focus on collective behavior, i.e.~coherent motion in the whole network or in large parts of it.  To this end, we study the eigenvalue and eigenvector statistics and identify significant sections in the motorway network. We find collective behavior in these significant sections and further explore its causes. We show that collectivity throughout the network cannot directly be related to the traffic states (free, synchronous and congested) in Kerner's three-phase theory. Hence, the degree of collectivity provides a new, complementary observable to characterize the motorway network.

\vspace{0.5cm}

\vspace{1cm}

%\pacs{ 89.65.Gh 89.75.Fb 05.10.Gg}

%\keywords{econophysics, complex systems, statistical analysis }
\noindent\rule{\textwidth}{1pt}
\vspace*{-1cm}
{\setlength{\parskip}{0pt plus 1pt} \tableofcontents}
\noindent\rule{\textwidth}{1pt}

\section{Introduction}
\label{sec1}

In complex systems~\cite{Ladyman2013,Ziemelis2001}, linear interactions between constituents can be visualized in the structures of correlation matrices. Due to non-stationarity~\cite{Schmitt2013,Stepanov2015,Wang2016} or a lack of sufficient data, the length of the time series for a meaningful analysis is limited. Resulting from finite time series, correlation matrices are to a large extent noise dressed. To distinguish significant information from noise, various methods are used. One such method is random matrix theory (RMT)~\cite{Guhr1998,Plerou2002,Potestio2009} that describes the bulk of eigenvalue spectra of noise-dressed correlation matrices~\cite{Laloux1999,Guhr2003}. The bulk of the spectra comprising of the small eigenvalues indicates the noise, i.e., pure randomness. In contrast, the larger eigenvalues outside the spectral bulk carry significant, non-random information. For example, it has been found that the largest eigenvalue manifests the market collective motion~\cite{Laloux2000,Plerou2002} in correlation matrices of financial time series corresponding to industrial sectors such as energy, information technology etc.~\cite{Gopikrishnan2001,Plerou2002,Wang2016a,Wang2016b,Benzaquen2017}. An analysis of the eigenvectors belonging to the large eigenvalues reveals the participating stocks and thus the companies~\cite{Gopikrishnan2001,Plerou2002}. 

A traffic network as a complex system features multiple traffic flows and velocities, measured by induction loops and averaged over 1-min intervals. According to the flow-density diagram~\cite{Kerner2004} in traffic systems, the same value of traffic flows may correspond to two different traffic states, i.e., a free and a congested traffic state. In contrast, the velocity-density diagram~\cite{Kerner2004} manifests that the velocity distinguishes the two traffic states well.  Depending on the conditions, for instance, road constructions, the weather or speed limitations, different road sections, i.e., the nodes in the network, have different (average) velocities. If the average velocity in a road section is lower than the velocity for maximal traffic flow~\cite{Kerner2004}, a congested traffic state appears, otherwise, a free traffic state is seen. In our analysis, we view the road sections as interacting constituents. This is formally similar to our previous analyses of stocks as interacting constituents in financial markets~\cite{Wang2016a,Wang2016b,Wang2018,Heckens2020}, but there are obvious differences: Road sections encode a geographical information, i.e, they represent a true topology. Furthermore, the non-stationarity contains seasonalities and periodicities, e.g., rush hours, and hence strong non-Markovian features.

Most previous studies on traffic networks are devoted to the modeling and simulation of traffic states either from a macroscopic or from a microscopic viewpoint~\cite{Nagel1992,Schadschneider1993,Lovaas1994,Schreckenberg1995,Hoogendoorn2001,Wong2002,Fellendorf2010,Treiber2013}. Due to the lack of available traffic data, there are less empirical studies~\cite{Kerner2002,Bertini2005,Schonhof2007,Kerner2004}. Moreover, first studies~\cite{Wang2020} indicate that the degree of complexity in traffic networks is very high, even compared to stock markets, partly due to the above mentioned reasons, partly because a variety of relevant time scales occurs. To unveil statistical characteristics and collective behavior, we transfer our previously developed methods~\cite{Guhr1998,Plerou2002} to the complete motorway network of North Rhine-Westphalia (NRW), the most populous state in Germany.

To figure out the behavior of road sections, we transfer the above procedures from a financial market to the complete motorway network in NRW. With plenty of traffic data, the interaction of motorway sections therefore can be uncovered by their cross-correlations.  We are particularly interested in collective effects, i.e.~in coherent behavior of the whole network or of large parts of it.  By extracting the spectral information from the cross-correlation matrices, we identify significant sections corresponding to the largest eigenvalue. We thereby uncover collective behavior in the motorway networks. Establishing the relation to the geographic regions and time periods, we further analyze the cause of collectivity.

The paper is organized as follows. We introduce the used dataset and the data processing in section~\ref{sec2}, and then describe the methods for data analyses in section~\ref{sec3}. In section~\ref{sec4}, we perform the analysis and extract the results. We present our conclusions in section~\ref{sec5}.

\section{Datasets}
\label{sec2}
 
\begin{figure}[htbp]
\begin{center}
\begin{minipage}[c]{0.56\textwidth}
\begin{overpic}[width=1\textwidth,fbox]{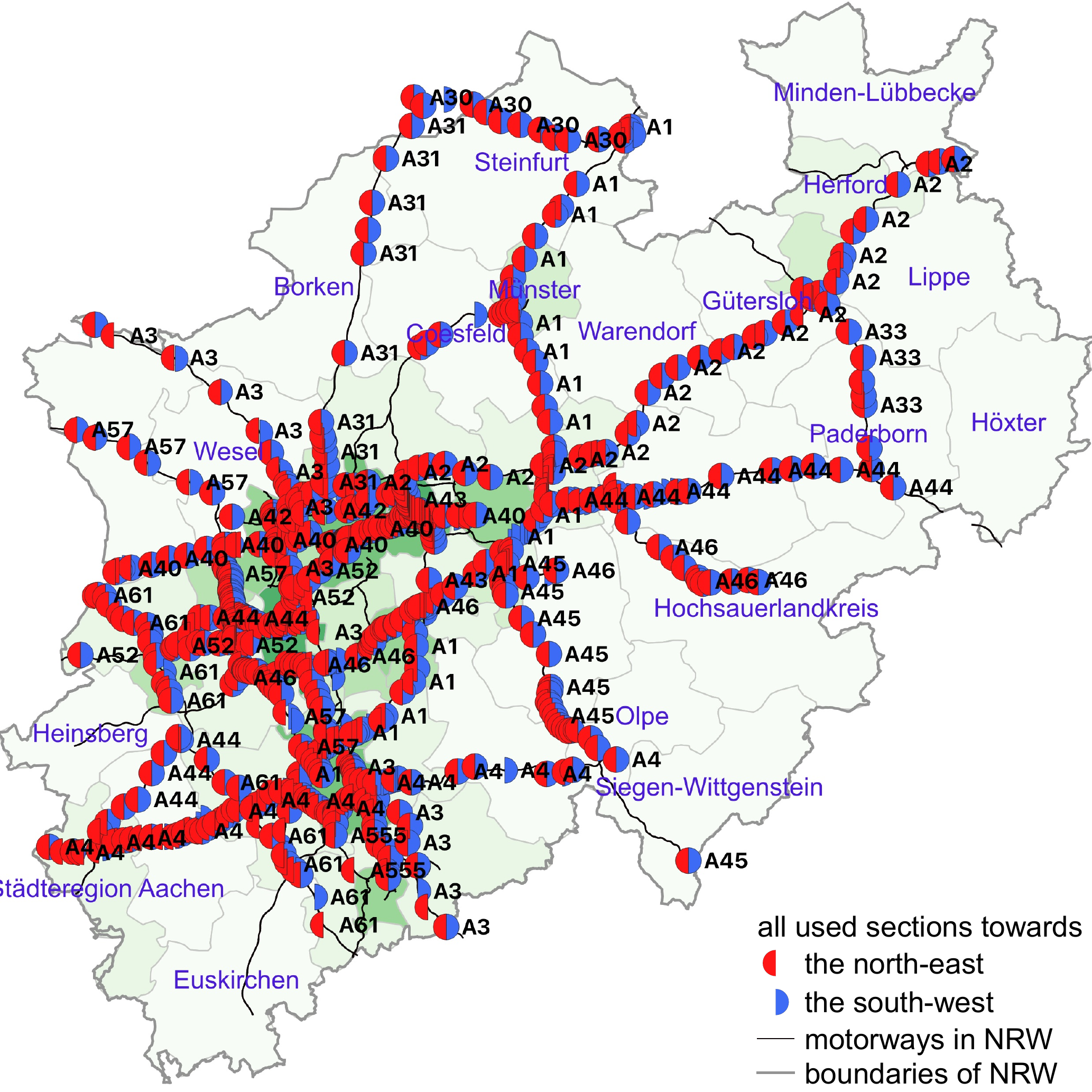}\put (1,93) {(a) whole region}\end{overpic} 
\begin{overpic}[width=1\textwidth,fbox]{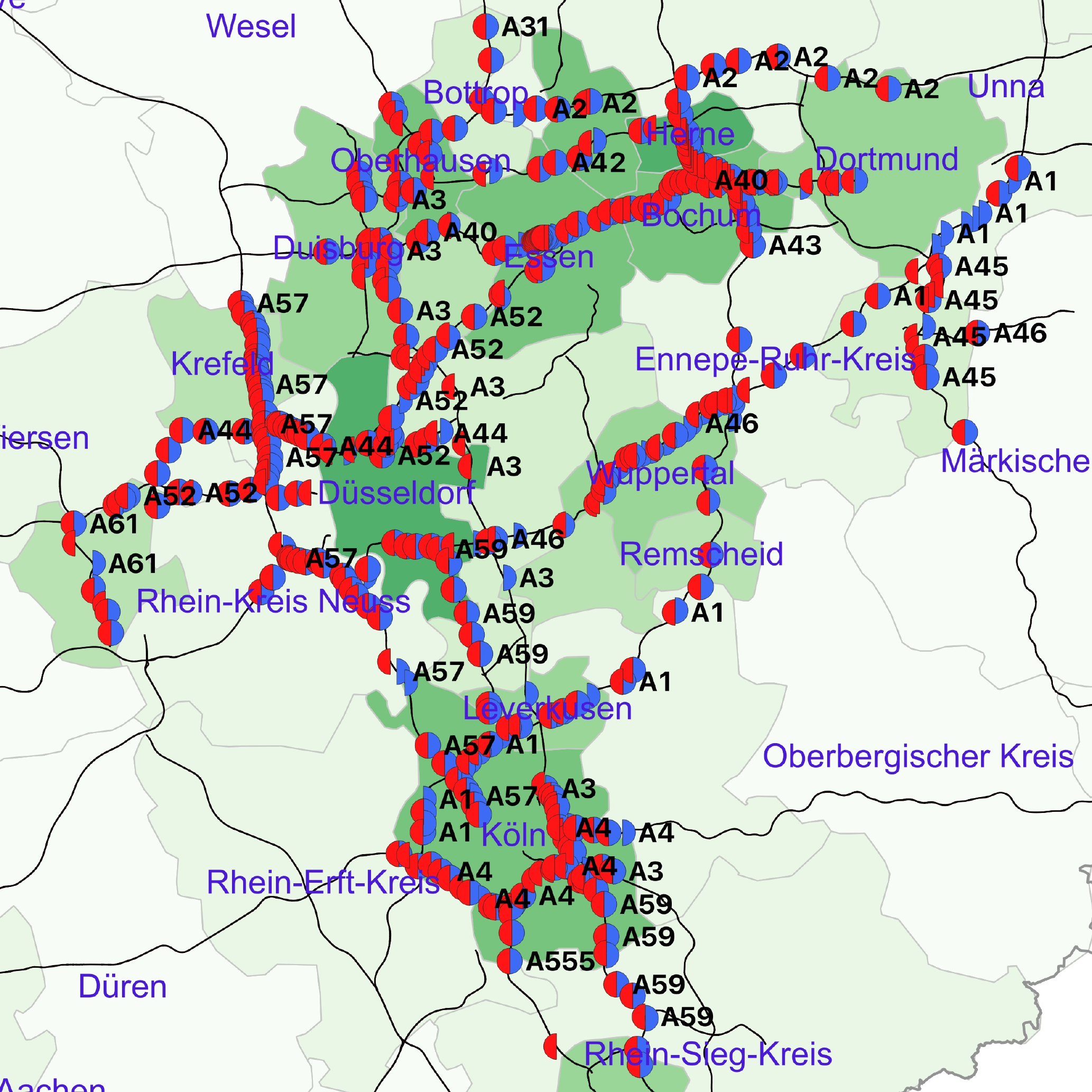}\put (1,92) {(b) central region}\end{overpic}
\begin{overpic}[width=1\textwidth,fbox]{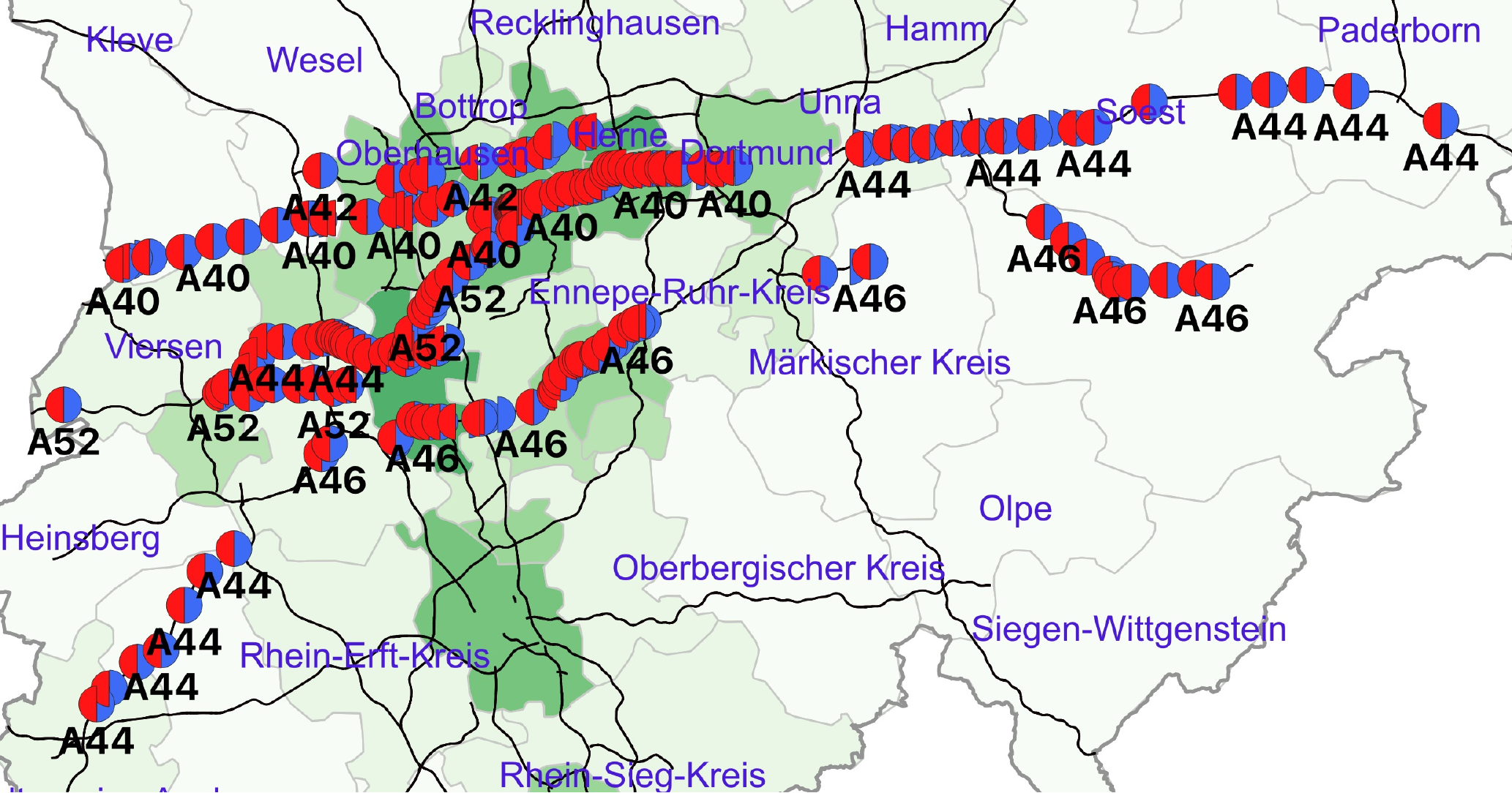}\put (1,46) {(c) parallel region}\end{overpic}
\end{minipage} \hspace*{0.8cm}
\begin{minipage}[c]{0.35\textwidth}
\caption{Three motorway networks in NRW: the complete network of the whole state (a), the central region with the high population density (b) and a region where the motorways are nearly parallel (c). The green background indicates the population density of districts in NRW. The darker the green background, the higher the population density. The data of administrative borders of districts (green lines) in NRW, licensed under BY-2.0, is provided by \copyright~GeoBasis-DE / BKG 2020~\cite{border,licence} and the data of population density in NRW, also licensed under BY-2.0, is provided by  \copyright~Statistische \"{A}mter des Bundes und der L\"{a}nder, Germany~\cite{population,licence}. The data of motorways (black lines) and the outside administrative boundaries of NRW (grey lines), licensed under ODbL v1.0, is provided by \copyright~OpenStreetMap contributors~\cite{osmcopyright, osm}. The map is developed with QGIS 3.4~\cite{qgis}.}
\label{fig1}
\end{minipage}
\end{center}
\end{figure}

This study uses the traffic data collected by inductive loop detectors on 22 motorways in NRW, Germany. We consider three cases of motorway networks in NRW: the motorways distributed in the whole state, in the central region with the highest population density and in a region where the motorways are nearly parallel, as shown in figure~\ref{fig1}. The three cases separately contain $N=1179$, 679 and 372 motorway sections. To avoid distortion of our results due to poor data quality, we limit the number of missing values to be less than $60\%$ of the total number in a whole day. Thus, high-quality traffic data in 80 days (including 64 workdays and 16 holidays) of 2017 is available for this study. 

Our data set has a resolution of one minute and includes information on time, traffic flow and velocity for every lane of each section. The traffic flow quantifies the number of vehicles per unit time. Dividing it by the velocity results in the flow density, which measures the number of vehicles per unit distance. For each section $n$ at each time $t$, we aggregate the traffic flows $q_{nl}(t)$ and the flow densities $\rho_{nl}(t)$ across all lanes $l$ as to obtain the velocity $v_n(t)$ of that section,
\begin{equation}
v_n(t)=\frac{\sum_l  q_{nl}(t)}{\sum_l \rho_{nl}(t)} \ .
\label{eq2.3}
\end{equation}
We distinguish the cases of workdays and holidays, where the holidays include the weekends and all public holidays in NRW.

\section{Methods}
\label{sec3}

Time $t$ is measured in units of minutes, $T$ is the total number of points in time. We organize the time series of velocities $v_n(t)$ of length $T$ for $N$ sections into the rows of a $N\times T$ data matrix $G$,
\begin{equation}
G=\left[\begin{array}{ccc}
v_1(1) & \cdots & v_1(T) \\
\vdots & \ddots & \vdots \\
v_N(1) & \cdots & v_N(T)
\end{array}\right] \ .
\label{eq3.1}
\end{equation}
In our study, $T=1440$ and $N=$1179, 679 and 372 for the cases of the whole region, the central region and the region of parallel motorways, respectively. Using the mean value
\begin{equation}
\mu_n=\frac{1}{T}\sum_{t=1}^{T} v_n(t) \ 
\label{eq3.2}
\end{equation}
and the standard deviation
\begin{equation}
\sigma_n= \sqrt{\frac{1}{T}\sum_{t=1}^{T}\big(v_n(t)-\mu_n\big)^2} \ 
\label{eq3.3}
\end{equation}
for each section $n=1,\cdots,N$, we normalize the elements in each row of $G$ to zero mean and unit standard deviation
\begin{equation}
M_n(t)=\frac{v_n(t)-\mu_n}{\sigma_n} \ .
\label{eq3.4}
\end{equation}
This results in a $N\times T$ normalized data matrix $M$ with elements $M_n(t)$ and further yields a real-symmetric $N\times N$ correlation matrix by
\begin{equation}
C=\frac{1}{T}MM^{\dag} \ ,
\label{eq3.5}
\end{equation}
where the $\dag$ indicates the transpose. To evaluate empirical correlation matrices, we fill the missing values in the original data matrix $G$ with the nearest existing values. This facilitates the spectral decomposition of correlation matrices, as the resulting correlation matrix is positive-definite and has non-negative eigenvalues. Ignoring the missing values in $G$, the correlation matrix is ill-defined and also has negative eigenvalues. Nevertheless, the structural difference of the two resulting correlation matrices is very slight.

For comparison with the empirical correlation matrices, we also use Random Matrix Theory (RMT) and simulate a random correlation matrix $R$ with a $N\times T$ random data matrix $A$ whose elements are normally distributed,
\begin{equation}
R=\frac{1}{T}AA^{\dag} \ .
\label{eq3.6}
\end{equation}
Each correlation matrix $X$ ($X=C$ or $X=R$) is then decomposed into a diagonal matrix $\Lambda$ of the eigenvalues $\Lambda_n$ and an orthogonal matrix $U$ by
\begin{equation}
X=U\Lambda U^{\dag} \ ,
\label{eq3.7}
\end{equation}
where the $n$-th column $U_n$ of $U$ is the eigenvector corresponding to $\Lambda_n$. 

The spectral density of a correlation matrix is given by 
\begin{equation}
\rho(\lambda)=\sum\limits_{k=1}^{N}\delta (\lambda-\Lambda_k) \ .
\label{eq3.10}
\end{equation}
We notice the normalization to the total eigenvalue number $N$. RMT provides results for a fully random correlation matrix. The Marchenko-Pastur distribution~\cite{Marchenko1967}
\begin{equation}
\rho_\mathrm{MP}(\lambda)=\frac{T}{2\pi}\frac{\sqrt{(\Lambda_+-\lambda)(\lambda-\Lambda_-)}}{\lambda} \ 
\label{eq3.11}
\end{equation}
is valid for large dimensions $N$. The maximal and minimal limiting values of the argument $\lambda$ are
\begin{equation}
\Lambda_{\pm}=1+\frac{N}{T}\pm2\sqrt{\frac{N}{T}}=\left(1\pm\sqrt{\frac{N}{T}}\right)^2 \ .
\label{eq3.12}
\end{equation}
A comparison of empirical eigenvalue distributions with RMT results helps us to assess the randomness of our empirical results in the bulk region, i.e., for small eigenvalues. We then also look at the large eigenvalues.

To identify the significant participants in an eigenvector $U_n$ with elements $U_{kn}$, we introduce the inverse participation ratio~\cite{Guhr1998}
\begin{equation}
I_n=\sum_{k=1}^{N} U_{kn}^4 \ .
\label{eq3.13}
\end{equation}
Its reciprocal reflects the number of significant participants, i.e., of those with the largest $1/I_n$ absolute values $|U_{kn}|$ of eigenvector components. For example, if all components are equal, $U_{kn}=1/\sqrt{N}$, we have $1/I_n=N$, while $U_{kn}=\delta_{k1}$, say, yields $1/I_n=1$. In the sequel, we will also use $1/(NI_n)$ to facilitate comparison of analyses for different numbers $N$ of sections.

\section{Results and analyses}
\label{sec4}

In section~\ref{sec41}, we compare the distributions of eigenvalues with the Marchenko-Pastur distribution. In section~\ref{sec42}, we reveal the features of significant participants for several largest eigenvalues. We then analyze the causes of collective behavior of significant sections encoded by the largest eigenvalue in section~\ref{sec43}.

\subsection{Distributions of eigenvalues}
\label{sec41}

\begin{figure}[tbp]
\begin{center}
\includegraphics[width=0.85\textwidth]{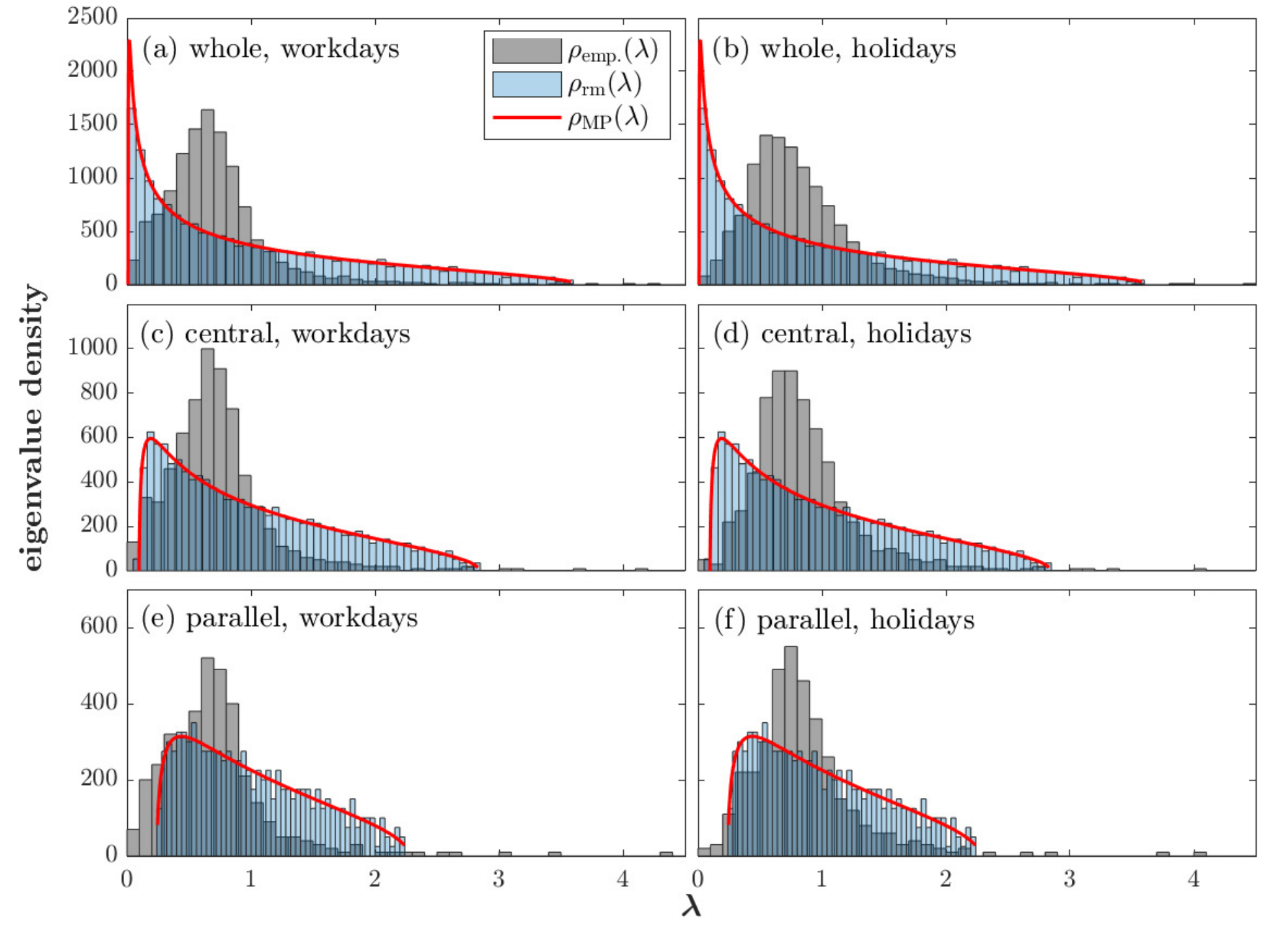}
\vspace*{-0.3cm}
\caption{Empirical eigenvalue density $\rho (\lambda)$ in the bulk region for the six cases, compared with the eigenvalue density of a purely random correlation matrix, $\rho_\mathrm{rm} (\lambda)$ simulated and Marchenko-Pastur distribution $\rho_\mathrm{MP} (\lambda)$.}
\label{fig2}
\end{center}
\vspace*{-0.3cm}
\end{figure}

The three networks in the different regions defined in section~\ref{sec2}, differentiating between workdays and holidays, yield six empirical correlation matrices $C$ in total. The probability densities of the smaller eigenvalues $\lambda$ are shown in figure~\ref{fig2}. For each case, the comparison of the empirical and the RMT distributions is shown. The simulated eigenvalue densities for fully random correlation matrices are fitted well by the Marchenko-Pastur distribution as expected. The small eigenvalues resulting from our empirical correlation matrices, however, depart evidently from the random ones. The large difference demonstrates the non-randomness of our empirical correlation matrices, indicating strong non-trivial correlation structures. Put differently, by looking at the largest eigenvalues, we can infer an important part of information encoded in these correlation matrices.

\begin{figure}[tbp]
\begin{center}
\includegraphics[width=0.85\textwidth]{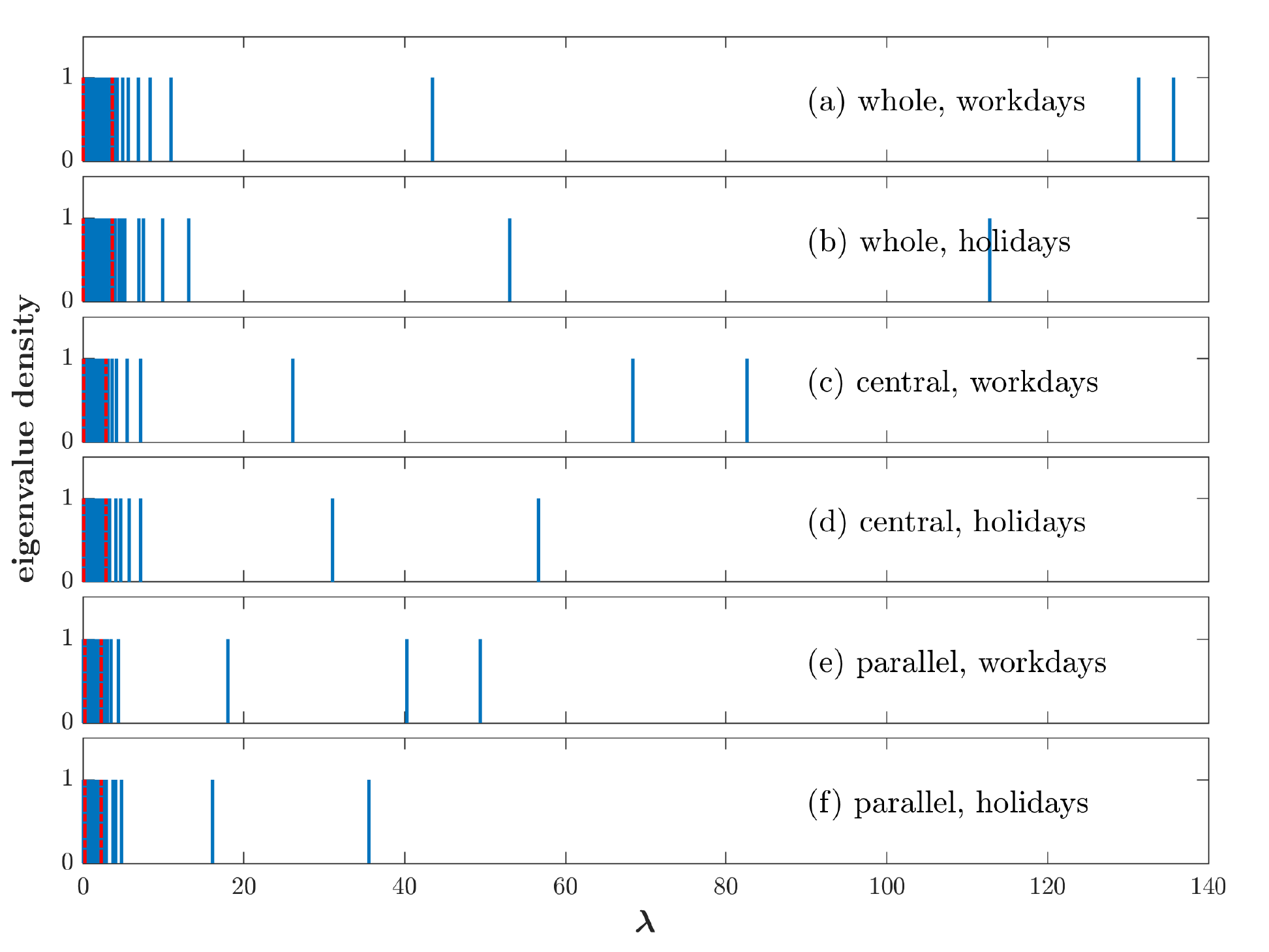}
\vspace*{-0.3cm}
\caption{Full empirical eigenvalue densities $\rho(\lambda)$ for the six cases, where the left and right red dash lines correspond to the minimal and maximal eigenvalues $\Lambda_\pm$ in the Marchenko-Pastur distribution $\rho_\mathrm{MP}(\lambda)$, respectively.}
\label{fig3}
\end{center}
\vspace*{-0.3cm}
\end{figure}

The full spectral densities are shown in figure~\ref{fig3}, which clearly reveals the larger (and small) eigenvalues outside the random regime. In each case, the largest one, two or three eigenvalues stand out. According to the analysis of financial markets, the largest eigenvalue in a correlation matrix is in proportion to the dimension of the correlation matrix, see a model discussion in reference~\cite{Guhr2003}. Its time dependence is closely related to the average of all correlation matrix elements~\cite{Plerou2002,Song2011,Stepanov2015,Pharasi2019}. This largest eigenvalue reflects the collective motion of the considered financial market. Hence, we may expect similar characteristics for our analysis of traffic networks, being aware of the systemic differences such as non-Markovian behavior. However, what kind of information carried by the largest eigenvalue is to be disclosed by examining the significant participants of motorway sections. 

\subsection{Significant participants}
\label{sec42}

The spectral decomposition or expansion of the correlation matrix $C$ is the following sum of dyadic matrices
\begin{equation}
C=\sum_{n=1}^{N}\Lambda_nU_n U_n^{\dag} \ ,
\label{eq4.1}
\end{equation}
where the elements, i.e., the correlation coefficients between the road sections, can be written as
\begin{equation}
C_{kl}=\sum_{n=1}^N\Lambda_nU_{kn} U_{ln}\ .
\label{eq4.2}
\end{equation}
Equation~\eqref{eq4.1} results in
\begin{equation}
CU_n=\Lambda_nU_n \ .
\label{eq4.3}
\end{equation}
Importantly, the eigenvalues are non-negative, and only the positive ones contribute in equations~\eqref{eq4.1} and~\eqref{eq4.2}. Thus the dyadic matrices $U_n U_n^{\dag}$ encode the information on positive or negative contributions to $C$ for a given eigenvalue $\Lambda_n$. More precisely, as the components $U_{kn}$ can be positive or negative, satisfying $-1\le U_{kn} \le +1$, the contribution to the correlation coefficient $C_{kl}$ between road sections $k$ and $l$ for a given eigenvalue $\Lambda_n$ is positive, if $U_{kn}$ and $U_{ln}$ have the same signs (positive or negative), but the contribution is negative, if they have opposite signs. Here, only the relative signs matter, not an overall sign: one easily sees that nothing changes in equations~\eqref{eq4.1} and~\eqref{eq4.2} when replacing $U_n$ with $-U_n$. The magnitude $|U_{kn}|$ quantifies the significance of this road section $k$ for a given eigenvalue $\Lambda_n$.  As shown, the eigenvalues strongly differ in magnitude, making the contributions just discussed more or less relevant. The larger the eigenvalue, the more relevant.

To find and quantitatively characterize the significant motorway sections for each case, we work out the reciprocal of the inverse participation ratio~\eqref{eq3.13} in the sequel, as it reveals the number of significant participants for the corresponding eigenvalue $\Lambda_n$. We then identify the motorway sections, corresponding to the largest $1/I_n$ absolute eigenvector components, as the significant participants for $\Lambda_n$.

\begin{figure}[tb]
\begin{center}
\includegraphics[width=1\textwidth]{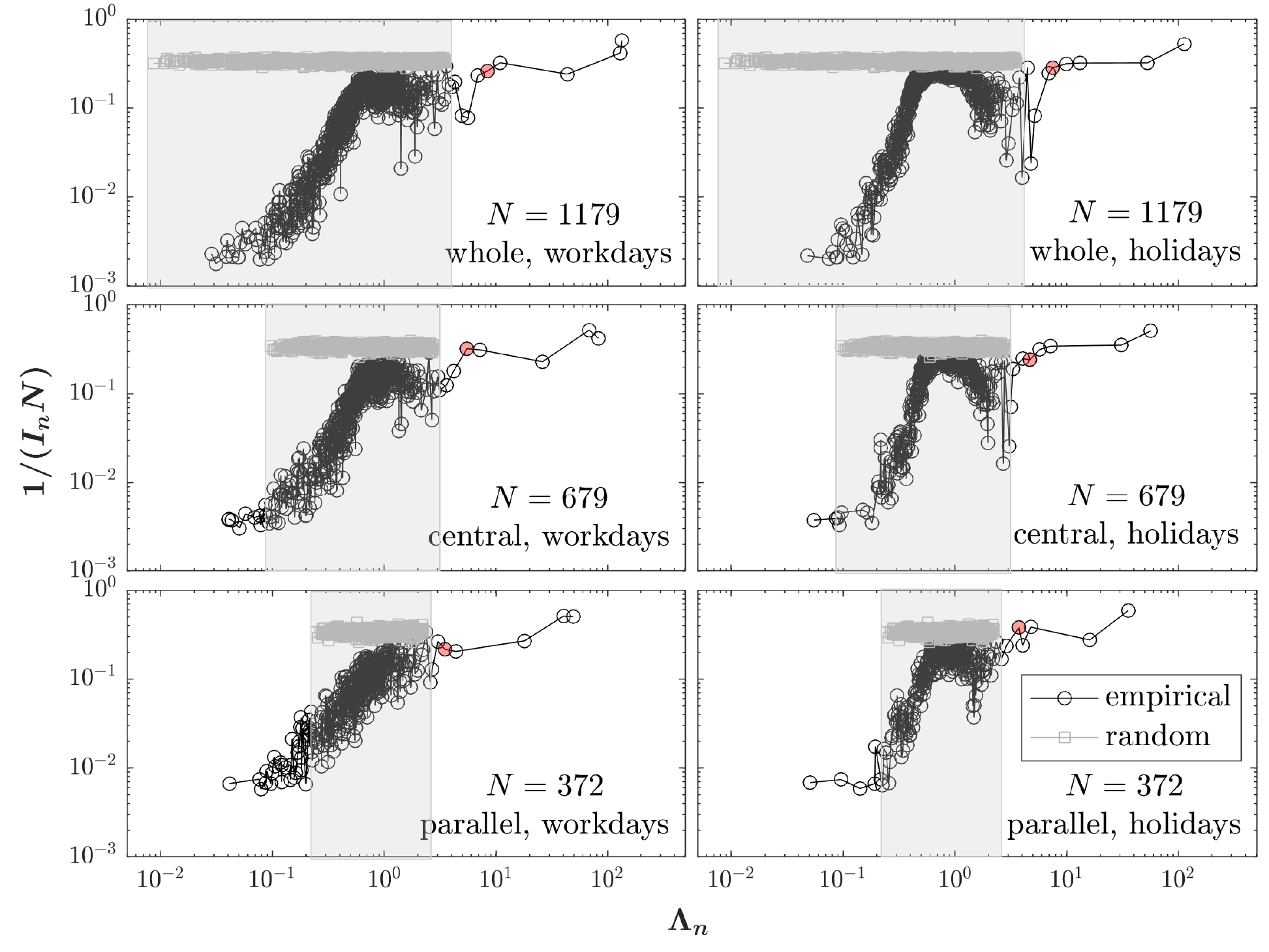}
\caption{Dependence of the ratio $1/(I_nN)$ of significant participants to all participants for the eigenvalues $\Lambda_n$ in the six cases. The markers filled with red color indicate the ratios of significant participants for the fifth eigenvalues. The gray areas correspond to the ranges of random eigenvalues.}
\label{fig4}
\end{center}
\end{figure}

\begin{figure}[htbp]
\begin{center}
\begin{minipage}[c]{0.56\textwidth}
\begin{overpic}[width=1\textwidth,fbox]{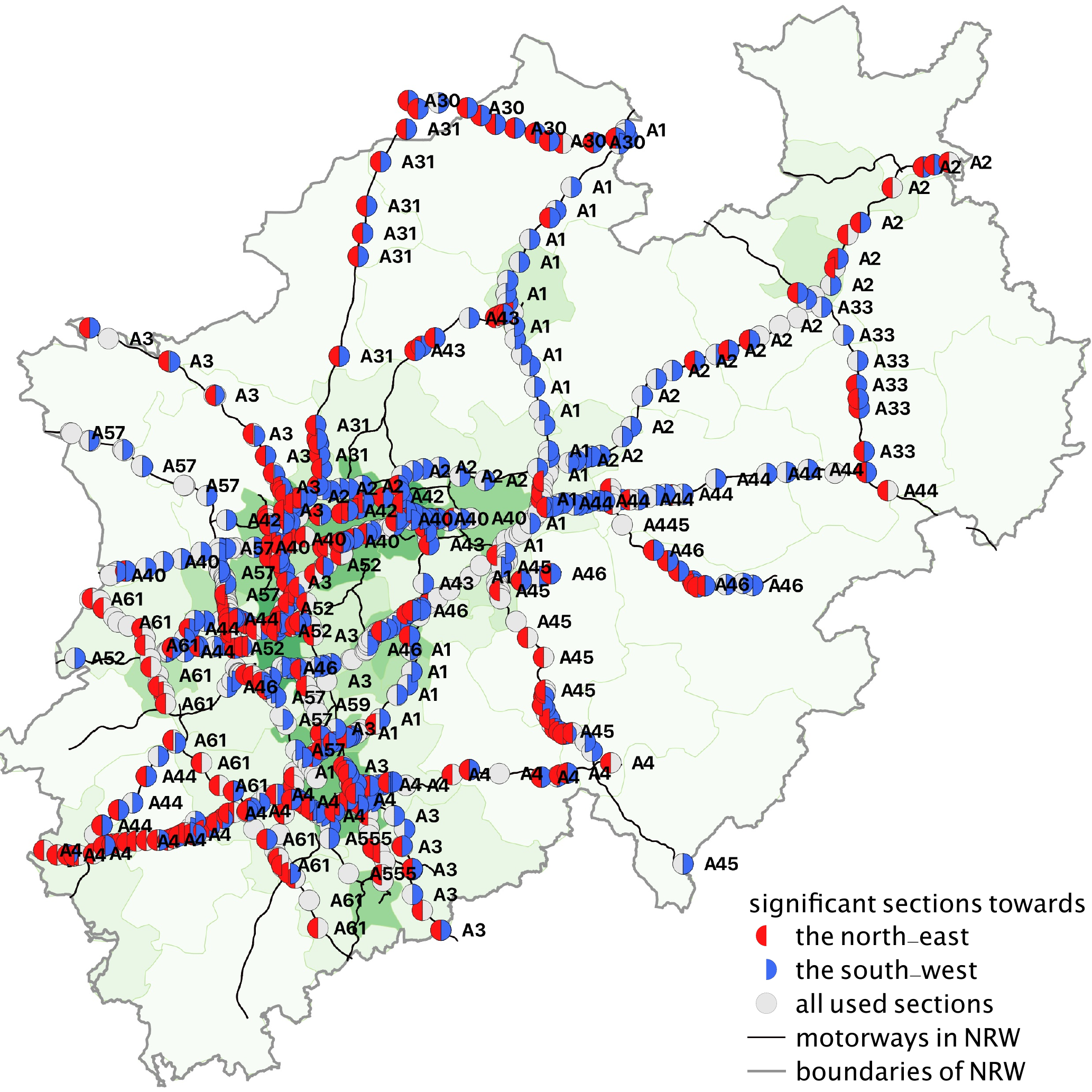}\put (1,93) {(a) whole region}\end{overpic} 
\begin{overpic}[width=1\textwidth,fbox]{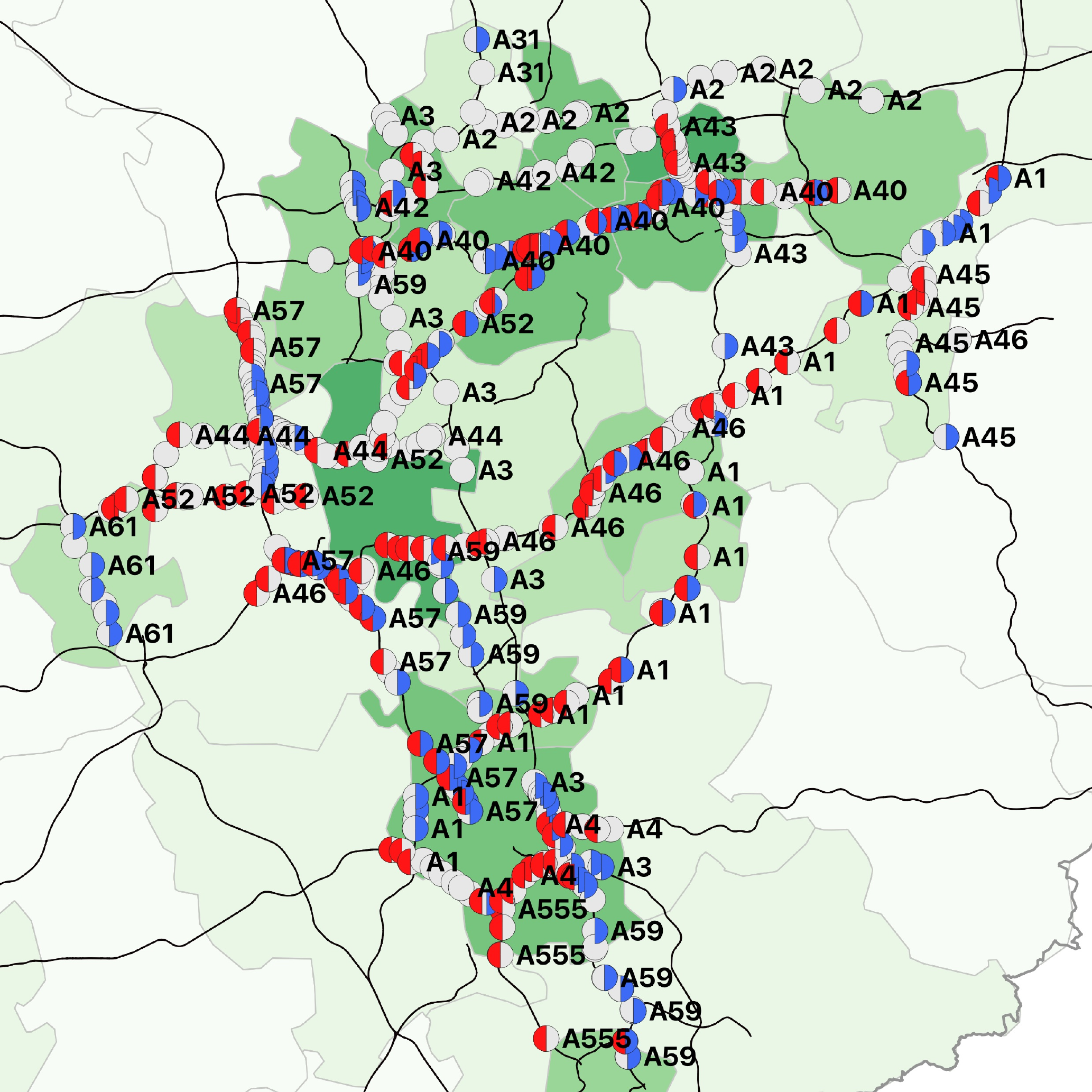}\put (1,93) {(b) central region}\end{overpic}
\begin{overpic}[width=1\textwidth,fbox]{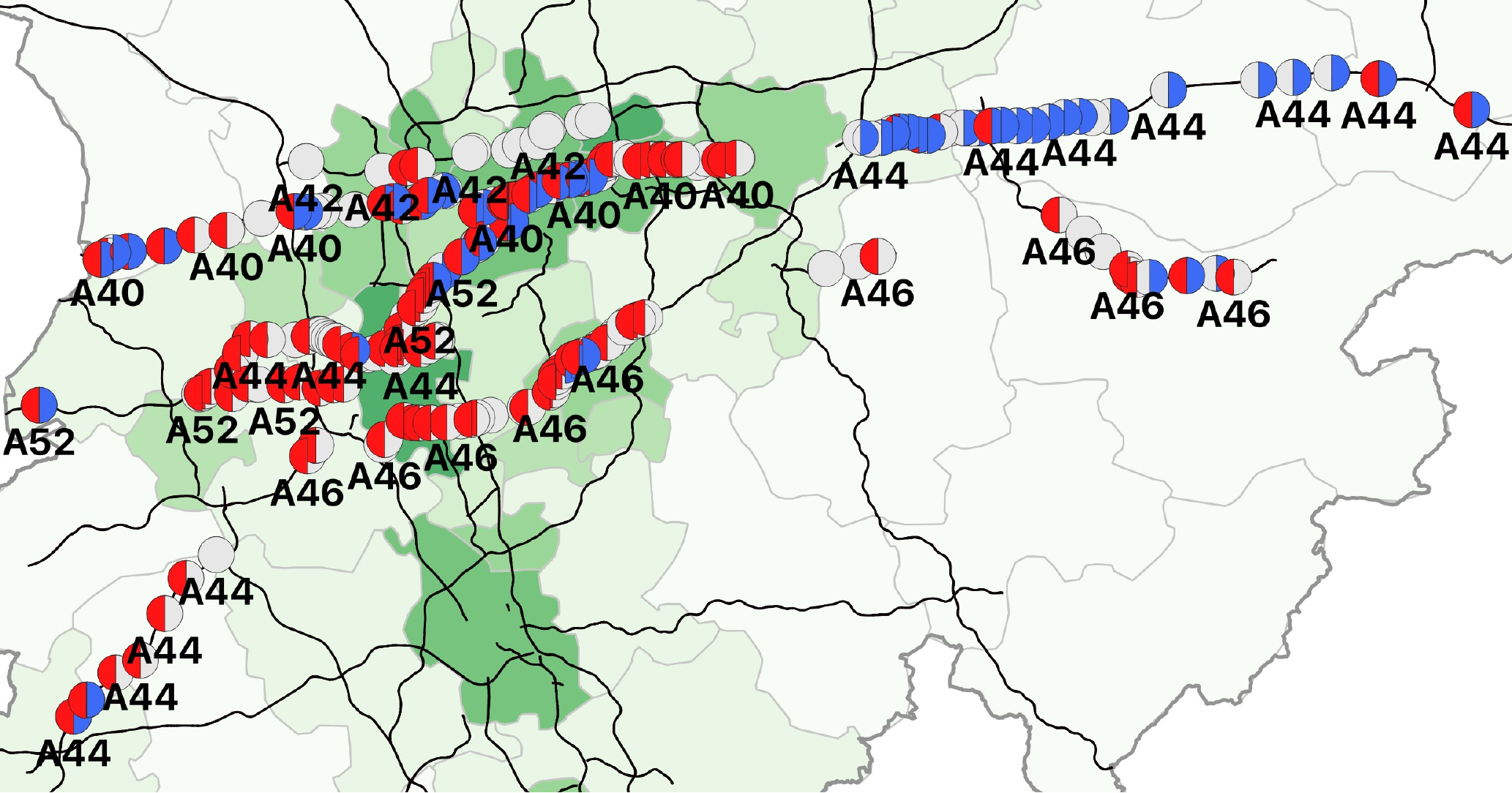}\put (1,47) {(c) parallel region}\end{overpic}
\end{minipage} \hspace*{0.8cm}
\begin{minipage}[c]{0.35\textwidth}
\caption{Geographic distributions of the significant sections for the largest eigenvalues in the three workday cases. For the information and the data source of the base map, refer to figure~\ref{fig1}. }
\label{fig5}
\end{minipage}
\end{center}
\end{figure}

\begin{figure}[htbp]
\begin{center}
\begin{minipage}[c]{0.56\textwidth}
\begin{overpic}[width=1\textwidth,fbox]{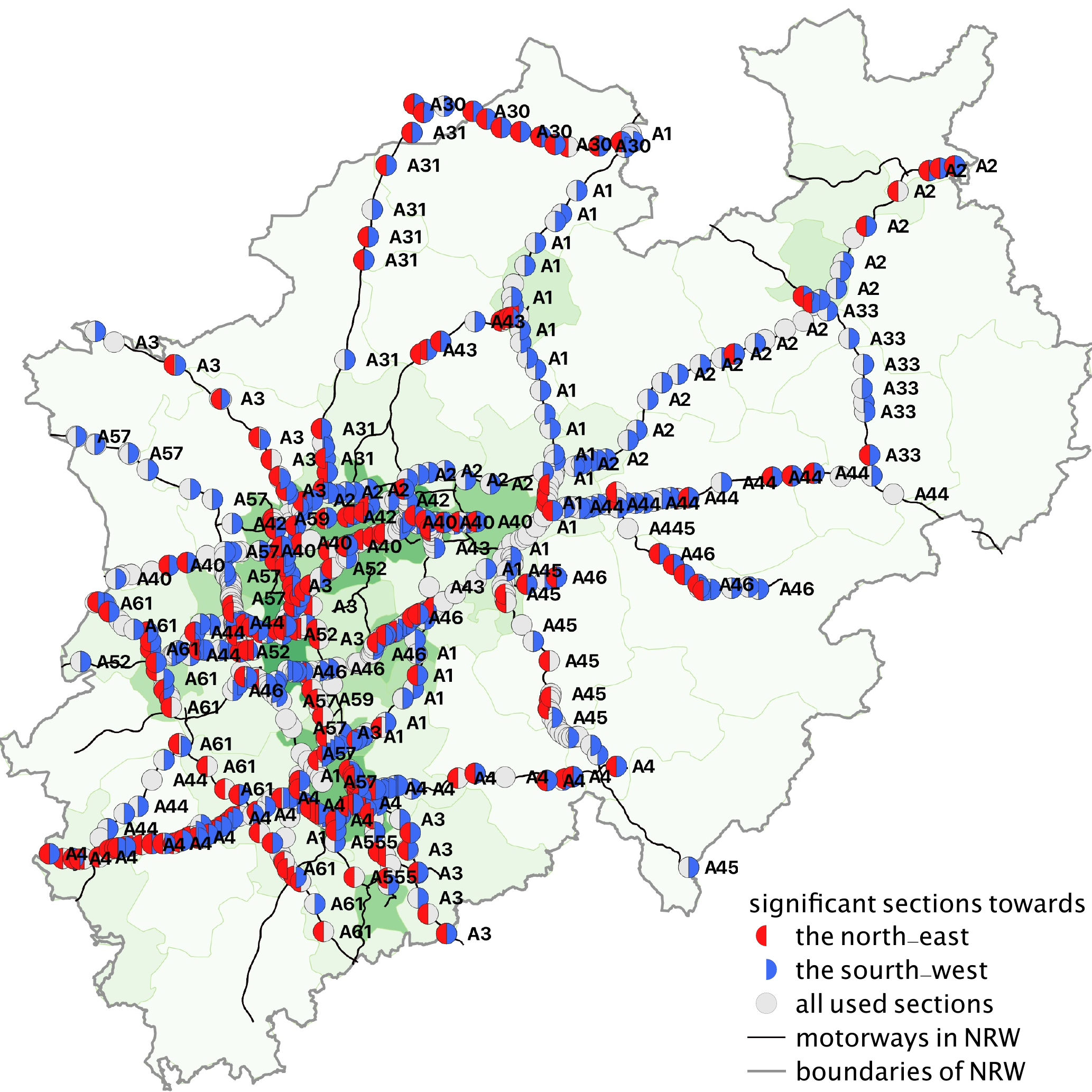}\put (1,93) {(a) whole region}\end{overpic} 
\begin{overpic}[width=1\textwidth,fbox]{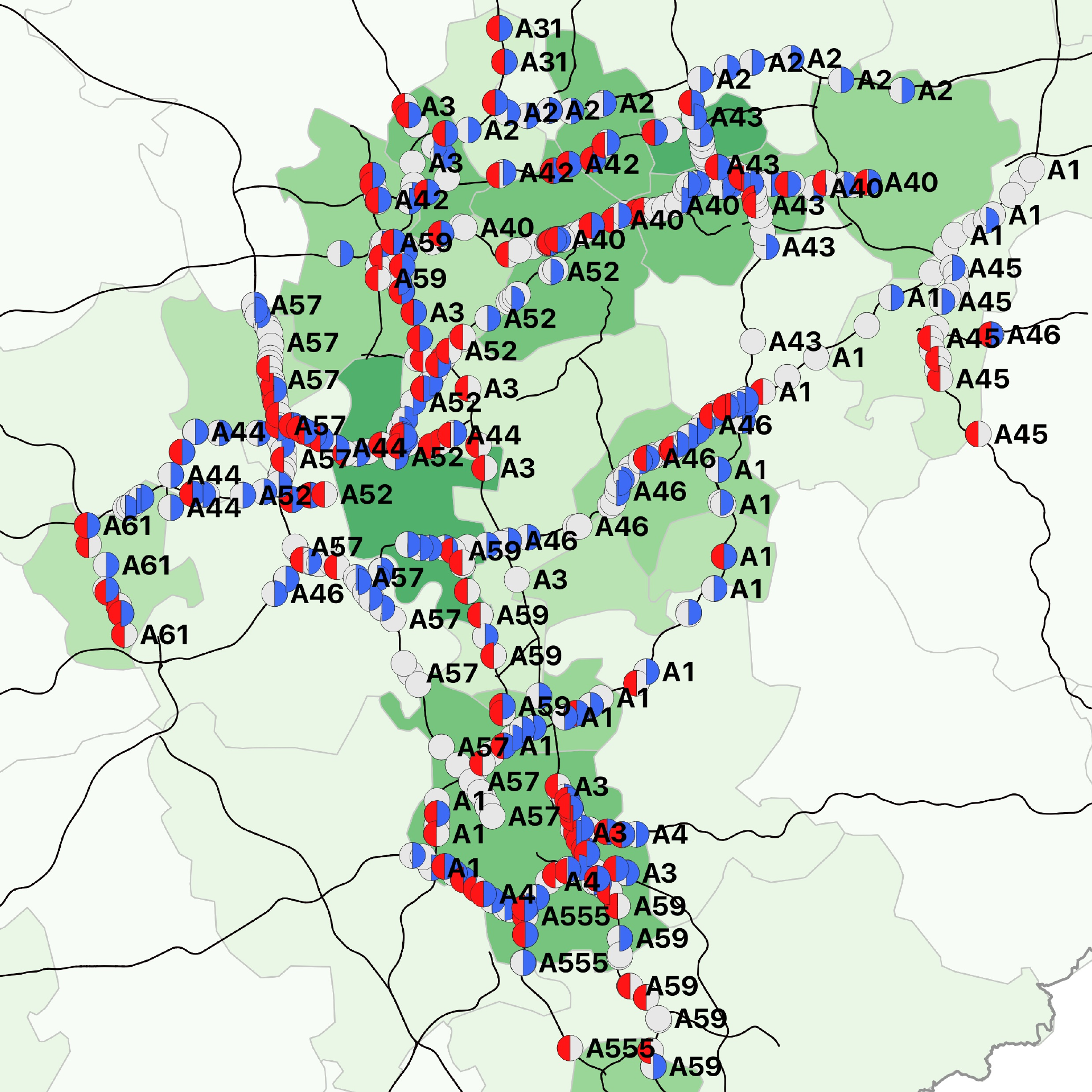}\put (1,93) {(b) central region}\end{overpic}
\begin{overpic}[width=1\textwidth,fbox]{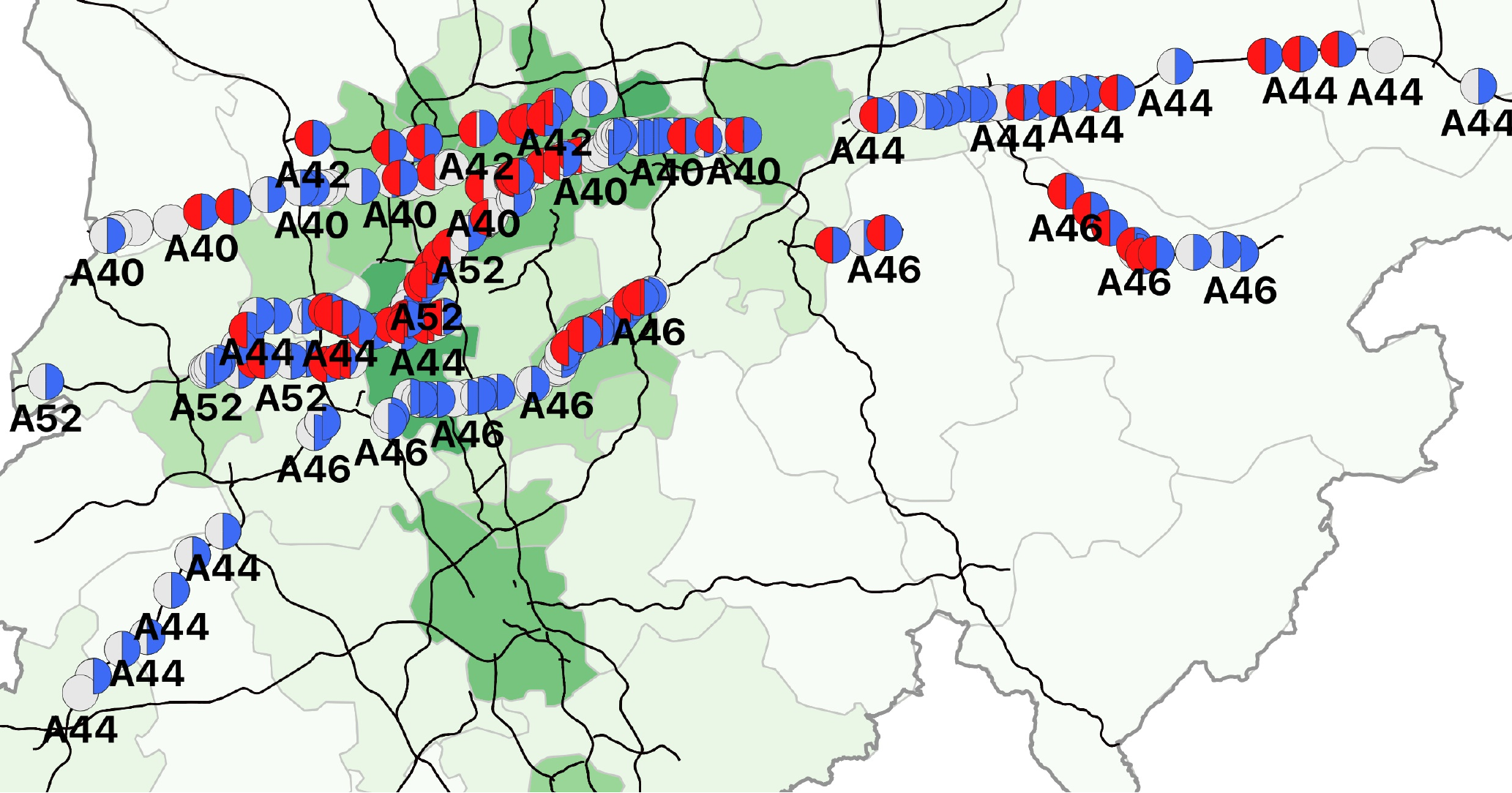}\put (1,47) {(c) parallel region}\end{overpic}
\end{minipage} \hspace*{0.8cm}
\begin{minipage}[c]{0.35\textwidth}
\caption{Geographic distributions of the significant sections for the largest eigenvalues in the three holiday cases. For the information and the data source of the base map, refer to figure~\ref{fig1}. }
\label{fig6}
\end{minipage}
\end{center}
\end{figure}

To account for the differences in the number $N$ of sections, we introduce the ratio $1/(I_nN)$ of significant participants to participants, i.e., sections. Figure~\ref{fig4} displays the results for all eigenvalues. This is in contrast to the random data which show the same ratio of significant participants for different eigenvalues. In particular, for the eigenvalues outside the range of randomness, the small ones have less significant sections than the large ones. Here, we focus on the large eigenvalues, especially the largest eigenvalue. We list the numbers and ratios of significant sections for the largest eigenvalue in Table~\ref{tab1}. In each case, around half of all sections, in contrast to almost all sections as in the case of financial markets~\cite{Plerou2002}, are found to be significant. The difference is simply due to the bidirectional structure of motorways. As shown in figures~\ref{fig5} and \ref{fig6}, the significant sections basically cover over all regions. Another interesting phenomenon is that the ratios of significant participants for the largest two eigenvalues are similar for workdays but differ largely for holidays. We highlight the circle corresponding to the fifth largest eigenvalue in figure~\ref{fig4}, as the ratios $1/(I_nN)$ of significant participants corresponding to this eigenvalue in the empirical and random cases are comparable in each subgraph.

\begin{table}[b]
\begin{footnotesize}
\begin{center}
\caption{Parameters for the first and the fifth largest eigenvalues in each case}
\begin{tabular*}{\textwidth}{c@{\hskip 0.4in}c@{\hskip 0.4in}c@{\hskip 0.4in}c@{\hskip 0.4in}c@{\hskip 0.4in}c@{\hskip 0.4in}c@{\hskip 0.4in}c}
\hlineB{2}
&&\multicolumn{2}{c}{for $\Lambda_\mathrm{max}$} & & \multicolumn{3}{c}{for $\Lambda_\mathrm{5th~max}$} \\
\cline{3-4}  \cline{6-8}
cases && $1/I_\mathrm{max}$ &ratio 	& & $p$-value    & $H_0$  & $H_1$ \\
\hline
whole region, workdays && 681	&57.76$\%$ &&  $3.00\times10^{-3}$ 	& reject    & accept	\\
whole region, holidays && 622 	&52.76$\%$ &&  0.15 				& accept  & reject  	\\
central region, workdays && 286 &41.83$\%$ && 0.55 				& accept  & reject 	\\
central region, holidays && 345  &50.81$\%$ &&  0.01 				& reject    & accept  	\\
parallel region, workdays &&  187 &50.27$\%$ &&  0.10				& accept  & reject 	 \\
parallel region, holidays && 218 &58.60$\%$ &&  0.66 				& accept  & reject 	 \\
\hlineB{2}
\label{tab1}
\end{tabular*}
\end{center}
\end{footnotesize}
\end{table}%

\begin{figure}[tbp]
\begin{center}
\includegraphics[angle=90, origin=c, width=1\textwidth]{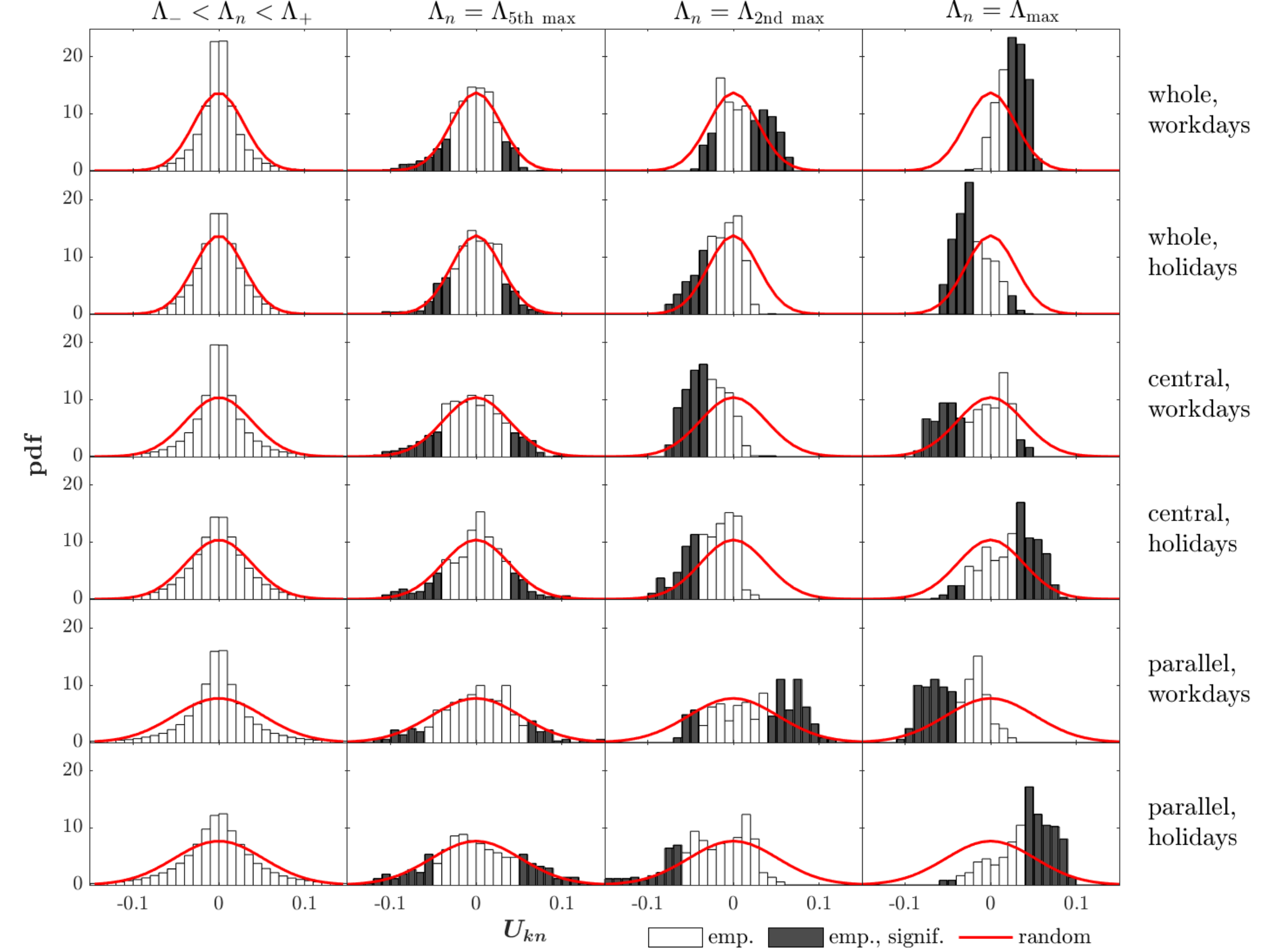}  
\caption{The distributions of components of eigenvectors corresponding to the bulks of small eigenvalues between $\Lambda_{-}$ and $\Lambda_{+}$, the fifth eigenvalues $\Lambda_\mathrm{5th~max}$, the second eigenvalues $\Lambda_\mathrm{2nd~max}$ and the largest eigenvalues $\Lambda_\mathrm{max}$ (shown in each column) in six cases (shown in each row). The empirical distributions of the significant participants are highlighted with black color. In each case, the empirical distribution of the bulk eigenvalues is fitted by a random distribution $\mathcal{N}(\mu_\mathrm{bulk},\sigma_\mathrm{bulk})$ with a red line. The other distributions in the same case are all compared with this distribution $\mathcal{N}(\mu_\mathrm{bulk},\sigma_\mathrm{bulk})$. The histograms are all normalized to one.}
\label{fig7}
\end{center}
\end{figure}

We take a closer look at the components of the eigenvectors that correspond to the first, the second, the fifth largest eigenvalues and to the bulk of eigenvalues inside the range of randomness, $\Lambda_{-}<\Lambda_n<\Lambda_{+}$, i.e., the grey regions in figure~\ref{fig4}. The probability densities are displayed in figure~\ref{fig7}. In each case, the histogram for the bulk of eigenvalues is fitted by a normal distribution. This is then compared with the histograms for the other three eigenvalues, where the eigenvector components of the significant participants are highlighted with black color. Due to the non-randomness as shown in figure~\ref{fig2}, the histogram of the eigenvector components for the bulk is fitted poorly by the normal distribution with mean value $\mu_\mathrm{bulk}$ and standard deviation $\sigma_\mathrm{bulk}$. In contrast, the normal distribution well describes the histogram of the components in the fifth eigenvector. Thus, the dyadic matrix $U_nU_n^\dagger$ for the fifth and all smaller eigenvalues are essentially matrices of numbers with absolute values smaller than one and random signs.  Thus, the corresponding contributions to the correlation coefficients $C_{kl}$ are random as well, positive or negative.  We notice that the significant components in the fifth eigenvector lie almost symmetrically in the positive and negative tails of the histogram. Hence taking into account all contributions for the fifth eigenvalue, say, and the smaller ones, there will also be cancelations of positive and negative contributions. This further explains the high importance of the larger eigenvalues for understanding the structure of the correlation matrix.

For the sake of completeness, we statistically confirm the randomness in the components of the fifth eigenvector indirectly by an Anderson-Darling test~\cite{Anderson1952}, a statistical test of whether a given sample of data follows a given probability distribution. When given a normal distribution, the Anderson-Darling test can detect the deviations of components in the fifth eigenvector from normality. Here, we set the null hypothesis, represented by $H_0$, that the components follow a normal distribution with mean value $\mu_5$ and standard deviation $\sigma_5$, and set an alternative hypothesis, represented by $H_1$, that the components are distributed non-normally. A $p$-value used in hypothesis testing is the evidence to support or reject the null hypothesis. A smaller $p$-value indicates a stronger evidence against the null hypothesis. If the $p$-value is lower than a preset significance level, the null hypothesis is rejected and invalid while the alternative hypothesis is accepted and valid. The $p$-values for the Anderson-Darling test are listed in Table~\ref{tab1}. At the $5\%$ significance level, the null hypotheses are accepted in four cases but rejected in the other two. The rejected null hypothesis implies that the fifth eigenvalue still carry some useful information. Since four out of six cases are validated to be randomly distributed and the other two approach random distribution by naked eyes, we use the histograms for the fifth eigenvector as an approximate benchmark for each case.

Although the ratios of significant sections for the largest two eigenvalues are similar, an obvious difference is revealed in the histograms of the eigenvector components. The histograms corresponding to the largest eigenvalue, in at least four cases, deviate stronger from the normal distribution than those corresponding to the second largest eigenvalue. In particular, the most significant participants for the largest eigenvalue lie on one side of the histogram, where the probability densities, i.e., the heights of bars in the histogram, for these sections are approximately uniform, especially in the cases of central and parallel regions. This feature is very different from the feature that most significant sections lie symmetrically on two tails of the histogram in our benchmark case for the fifth eigenvalues. It implies that a common force drives the most significant participants with the same bias, leading to a collective behavior in each case.

\begin{figure}[tbp]
\begin{center}
\includegraphics[angle=90, origin=c, width=1\textwidth]{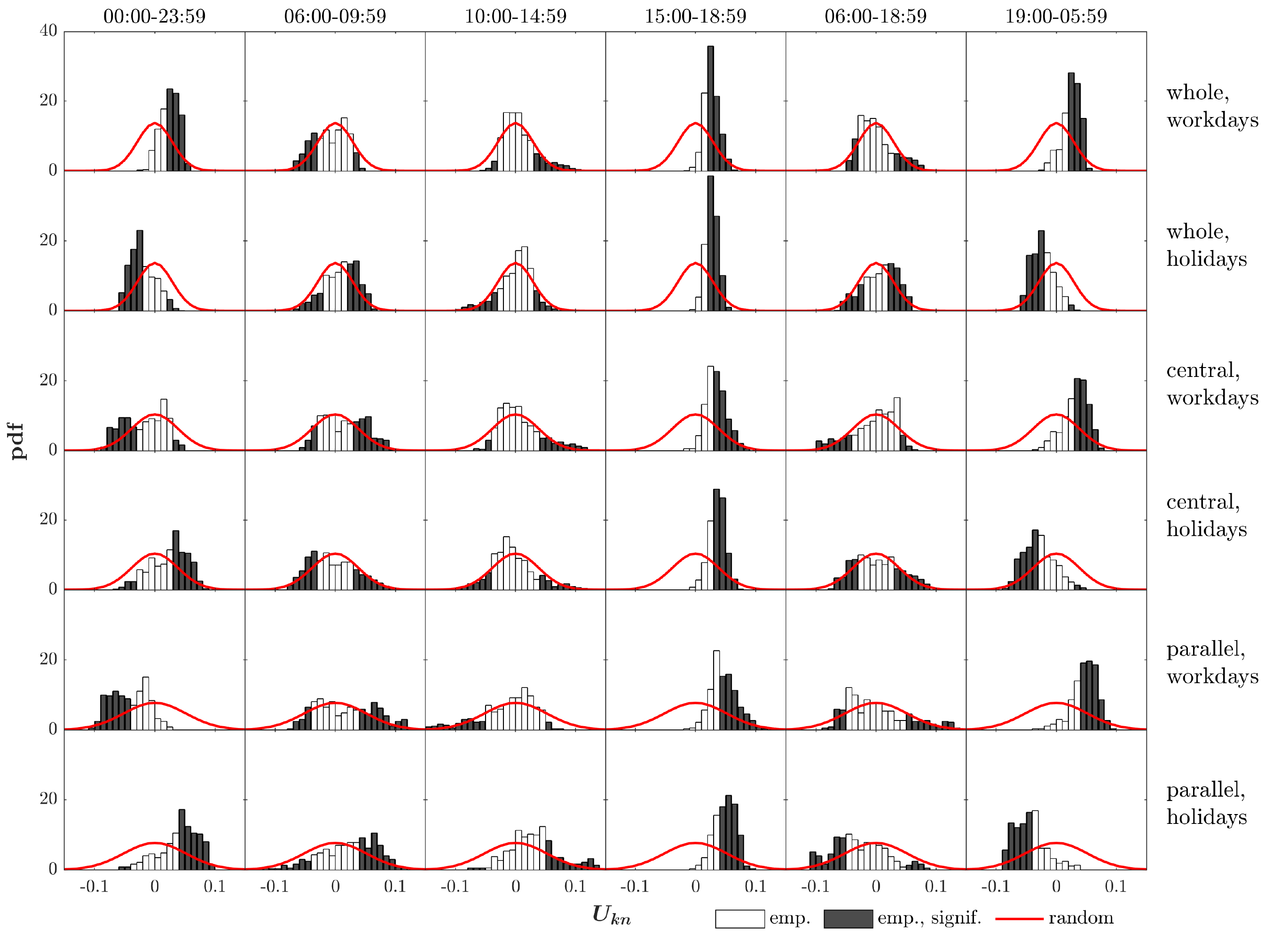}
\caption{The distributions of components of eigenvectors corresponding to the largest eigenvalues during different time periods (shown in each column) in one day. The empirical distributions of the significant participants are highlighted with black color. In each case (shown in each row), the empirical distribution are all compared with the random distribution $\mathcal{N}(\mu_\mathrm{bulk},\sigma_\mathrm{bulk})$. The histograms are all normalized to one.}
\label{fig8}
\end{center}
\end{figure}

\begin{figure}[tbp]
\begin{center}
\includegraphics[angle=90, origin=c, width=0.98\textwidth]{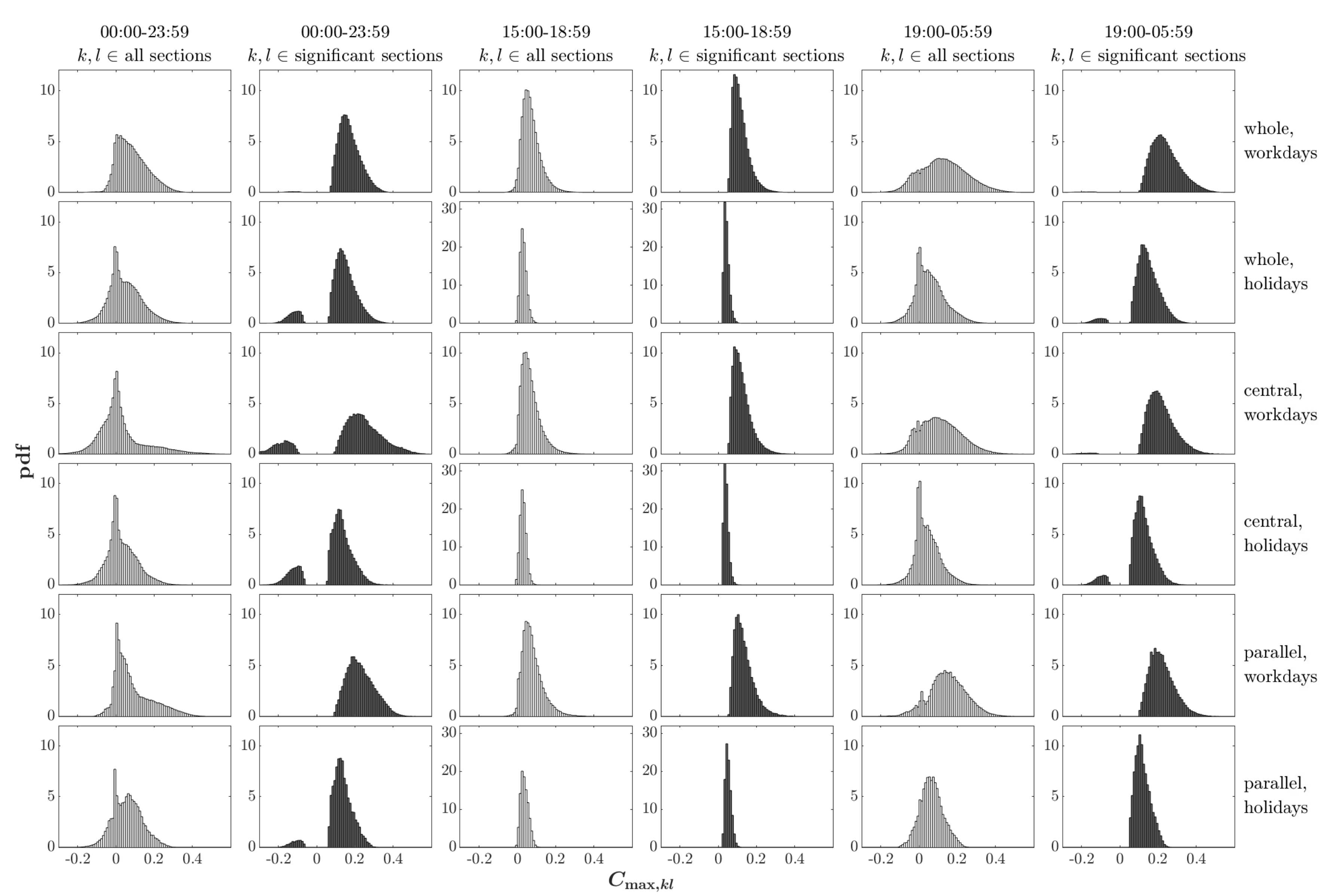}
\caption{The distributions of correlation components $C_{\mathrm{max},kl}$ corresponding to the largest eigenvalues for six cases (shown in each row) during three time periods (shown in each column) in one day.}
\label{fig9}
\end{center}
\end{figure}

\subsection{Causes of collective behavior}
\label{sec43}

To unveil the reason for the above-mentioned collective behavior in the significant sections, we separate each day into several time periods. These are the morning rush hours 06:00--09:59, the afternoon rush hours 15:00--18:59, the period between the two rush hours 10:00--14:59, the period of day time 06:00--18:59 and the period of night time 19:00--05:59. During each time period, we evaluate the correlation matrices of velocities for the six cases and then apply spectral decomposition to these matrices. Figure~\ref{fig8} shows the distribution of the eigenvector components corresponding to the largest eigenvalue for each time period of each case, where the distribution for a whole day 0:00--23:59 is displayed for comparison.

As seen in figure~\ref{fig8}, the collective behavior of significant sections is present during afternoon rush hours and during night time, respectively, for all cases, but is absent during other time periods. Here we notice that most of the significant sections lie either in the left or in the right tails of the distributions. During night time as well as during the whole day, we observe that the locations of the significant sections flip from the left to the right tails and vice versa. This phenomenon occurs because the positive and negative eigenvector components corresponding to the significant sections may contribute differently to the correlations, implying that the resulting contributions from significant sections can cancel each other out. To clarify that, we closer inspect the correlation coefficients
\begin{equation}
C_{\mathrm{max},kl} = \Lambda_{\mathrm{max}}U_{k,\mathrm{max}}U_{l,\mathrm{max}} 
\label{eq4.4}
\end{equation}
corresponding to the largest eigenvalue. In figure~\ref{fig9} we show the distributions of these correlations coefficients~\eqref{eq4.4} for all sections and only for the significant ones. As seen, the location flip of the contribution from significant sections in figure~\ref{fig8} has little impact on the distribution of the correlation coefficients. As long as both,  $U_{k,\mathrm{max}}$ and $U_{l,\mathrm{max}}$, have the same sign,  they contribute positively to the correlations coefficients  $C_{\mathrm{max},kl}$, i.e.~they lie on the positive side. This is so  for all cases during afternoon rush hours. For opposite signs,  contributions on the negative sides are found, even though they are  small, see the cases in the central region during a whole day. In  particular, with the same sign of eigenvector components, the  correlation distributions reveal heavier right tails for workdays  than for holidays.

\begin{figure}[tb]
\begin{center}
\includegraphics[width=1\textwidth]{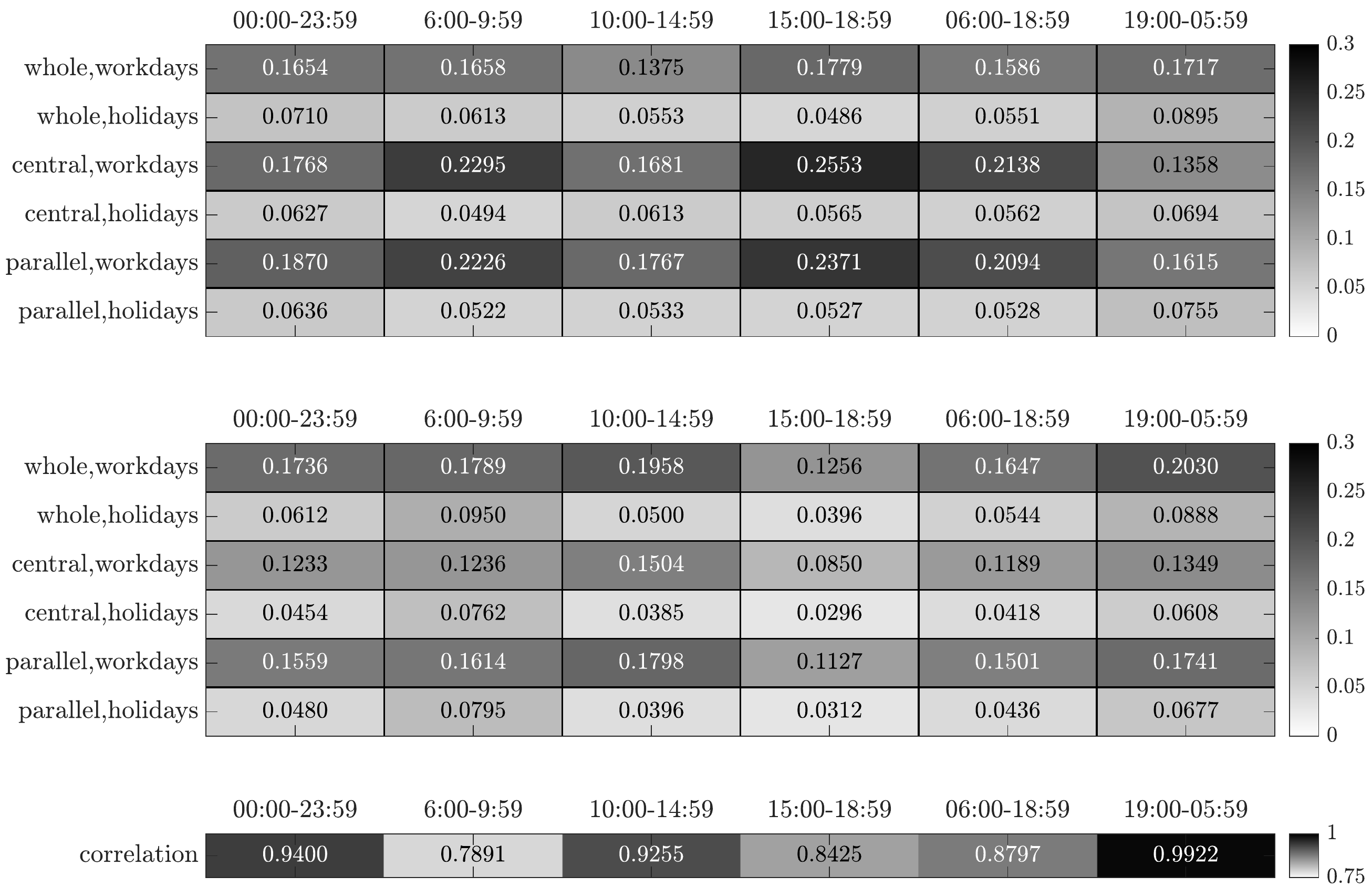}
\caption{Heat maps that visualize values by color in two-dimensional matrices, where the visualized values are the ratios of the numbers of the congested and all traffic states (top) and the ratios of truck flows to total flows (middle), respectively, for each case during each time period, and the correlations between two kinds of ratios across case series for each time period (bottom). The variation of color indicates the change of values. The cases are labeled at the left side of each row (top and middle) and  the time periods are labeled at the top of each column.}
\label{fig10}
\end{center}
\end{figure}

Furthermore, we now relate the collective behavior and its difference between workdays and holidays to traffic states and vehicle types. To this end, we use the ratios of the numbers of the congested and all traffic states as well as the ratios of truck flows to total flows, respectively, for each case during each time period, shown in figure~\ref{fig10}. We emphasize that a congested traffic state does not mean zero velocity for vehicles, rather, a lower velocity than the minimal velocity that the maximal traffic flow can hold on one section~\cite{Kerner2004}. Figure~\ref{fig10} shows the congestion ratios during morning and afternoon rush hours for each case. The ratios are significantly higher for the central and parallel regions on workdays than for other cases during the same time periods. Obviously, the striking collective behavior during afternoon rush hours in figure~\ref{fig8} is less likely to be related to a specific traffic state. Looking at the ratios of truck flows, the minimal value for each case is always found during afternoon rush hours. The low ratios of truck flows implies that the car flows highly dominate a traffic system. Comparing with a specific traffic state, the cars dominating the traffic system are highly responsible for the significance of sections during afternoon rush hours.

During night time, in view of the very high correlation between the two kinds of ratios, the truck flow due to speed limitations is very likely to cause the congestion. In contrast to holiday nights, the higher truck flows during workday nights produce higher congestion ratios on motorways, which enhance correlations between velocities. As a results, the distributions of the correlations coefficients $C_{\mathrm{max},kl}$ in figure~\ref{fig9} show heavier right tails for workday nights than for holiday nights. This line of reasoning also applies to the cases during afternoon rush hours on workdays and on holidays.

Nevertheless, comparing with the values for central and parallel regions during rush hours, the ratios of congestion during night time are still low enough to indicate a free traffic state, which dominates the motorways during night time and contribute considerably to the correlations of velocities among sections. In contrast to the collective behavior during afternoon rush hours, the collective behavior during a whole day is closer to the one during night time. One of the reasons is revealed by the similarity of the corresponding distributions in figures~\ref{fig8} and \ref{fig9}. Besides, the similarity is also visible in the ratios of congestion and the ratios of truck flows. This means the collective behavior during a whole day is very likely to share the same cause for the collective behavior during night time, i.e., free traffic states.

\begin{figure}[htbp]
\begin{center}
\includegraphics[width=1\textwidth]{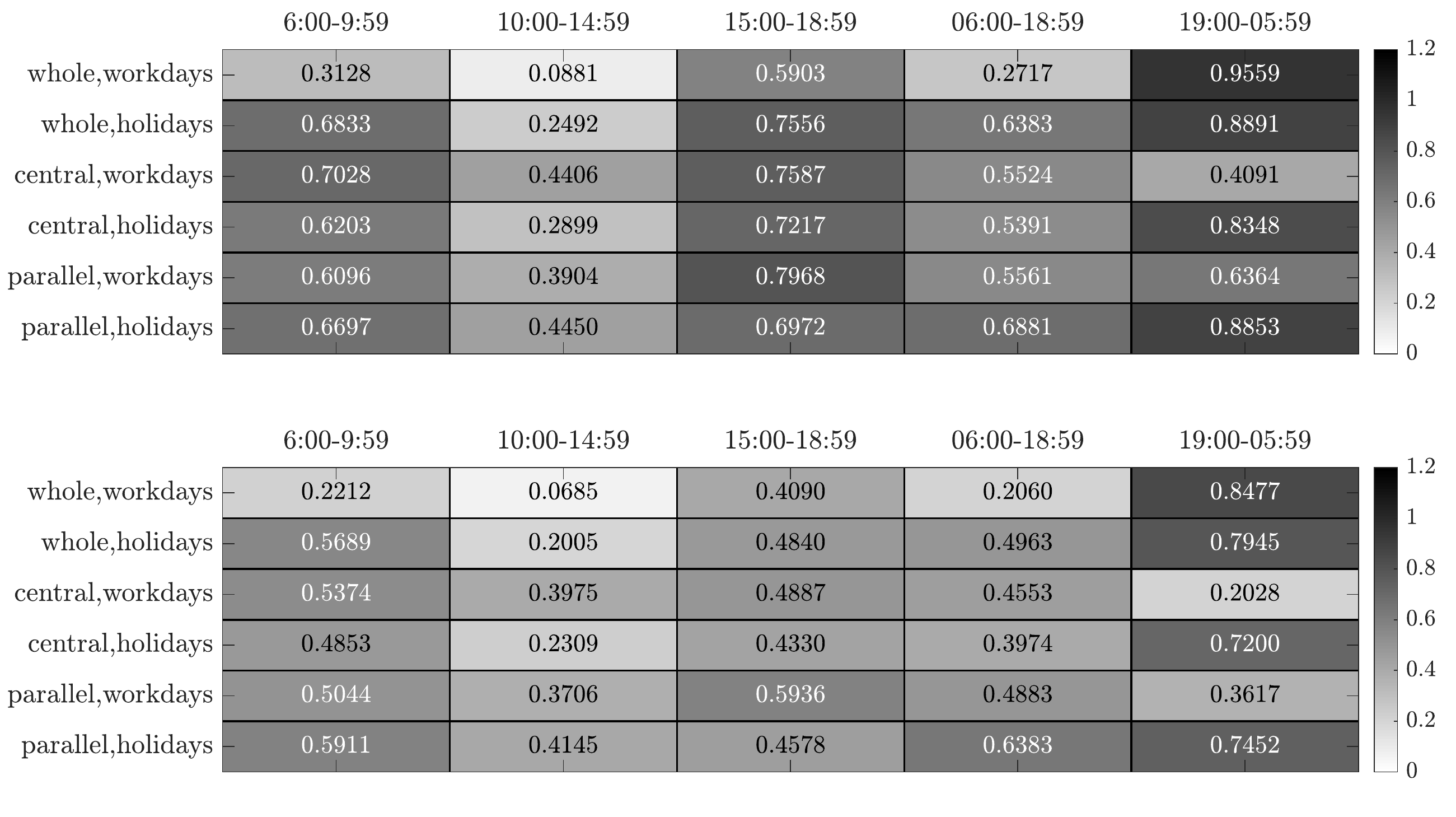}
\caption{Heat maps that visualize the overlap ratios by color in two-dimensional matrices. The variation of color indicates the change of overlap ratios. The overlap ratios on the top heat map is defined by $r_{xt}^{(1)}$ and in the bottom heat map by $r_{xt}^{(2)}$, where $x$ indicates the cases labeled at the left side of each row and $t$ indicates the time periods labeled at the top of each column.  }
\label{fig11}
\end{center}
\end{figure}

The above qualitative reasoning for the cause of collective behavior is supported by the following quantitative analysis. Let $S_{x}$ and $S_{xt}$ be the sets of the significant sections in case $x$ during a whole day and during a time period $t$, respectively. An overlap between the two sets can be written as $S_{x} \cap S_{xt}$, while a combination between the two sets as $S_{x} \cup S_{xt}$. We use the notation $| S |$ for the number of sections in a set $S$. We define two overlap ratios of the significant sections during a whole day with the ones during a given time period as
\begin{equation}
r_{xt}^{(1)}=\frac{|S_{x} \cap S_{xt}|}{|S_{x}|} \quad \mathrm{and} \quad r_{xt}^{(2)}=\frac{|S_{x} \cap S_{xt}|}{|S_{x} \cup S_{xt}|} \ ,
\end{equation}
respectively. Here, $r_{xt}^{(1)}$ measures the extent that the significance of the sections in the set $S_{x}$ is caused by the traffic states during the time period $t$, which may also cause the significance of the sections not contained in $S_{x}$. By including the non-overlapping significant sections from $S_{xt}$ in the denominator, $r_{xt}^{(2)}$ measures the ratio of the common sections whose significance is caused by traffic states during $t$ to all distinct sections in both sets $S_{x}$ and $S_{xt}$. The two kinds of overlap ratios are shown in figure~\ref{fig11}. As for $r_{xt}^{(1)}$, a majority of the significant sections for a whole day has high overlap with the significant sections during morning rush hours, afternoon rush hours and night time. Hence, a common cause for the same significances is present in a whole day and in the other three time periods. As for $r_{xt}^{(2)}$, the high overlap ratios are visible only during night time. This implies that the cause for the significance of sections during a whole day is close to the one during night time where the central and parallel regions behave differently on workdays. Moreover, it also suggests that a different cause for the significance of sections during the two rush hours is present. As the free traffic state dominates during night time, the collective behavior in the significant sections mainly originates from the free traffic state in the whole region during a whole day with a probability of at least $79\%$, and in the central and parallel regions during a whole holiday with a probability of at least $72\%$. Regarding the relative high overlap ratios with the significant sections during periods of two rush hours, the collective behavior in the central and parallel regions during a workday is more likely to result from the congested traffic state.

\section{Conclusions}
\label{sec5}

We studied the behavior in significant sections of three networks, the complete motorway network, the central motorways and the parallel motorways of NRW, differentiating between workdays and holidays. In altogether six cases, we performed a spectral decomposition for the correlation matrices of velocities measured over a whole day. The distributions of the resulting small eigenvalues deviate remarkably from the Marchenko-Pastur distribution. This and the presence of large, isolated eigenvalues show that the empirical correlation matrices are non-random and contain strong non-trivial correlation structures.

With the help of inverse participation ratios, we determined the ratio of significant sections to all sections for each eigenvalue. In most cases, the largest eigenvalue holds the highest ratio, which is around $50\%$ due to the bidirectional structure of motorways. For the largest eigenvalue, we further identified the significant sections through the components of the corresponding eigenvector. The significant sections almost cover all motorways on the geographic map of each region. Besides, for each case, most significant sections lie on one side of the distribution of eigenvector components corresponding to the largest eigenvalue, suggesting that a common force drives these sections with the same bias. We therefore concluded that the largest eigenvalue reveals a collective behavior of the significant sections.

By exploring the causes for the collective behavior in terms of traffic states, we found that the collective behavior for the whole motorway network during a whole day and for the central and the parallel motorways during a holiday originate from the free traffic state, while the collective behavior for the central and the parallel motorways during a workday very likely comes from the congested traffic state. We also found that the collective behavior of significant sections is absent during some short time periods, such as morning rush hours, but is remarkably present during afternoon rush hours in all cases. Hence, collectivity is a new, complementary measure and observable to characterize traffic networks. While the states in Kerner's three-phase traffic theory are defined and found in a linear context, i.e.~when looking at motorway sections in their topological sequence, collectivity analyzed in the correlation structure adds information on the network as such and on its interdependencies and mutual influences.

Collective behavior in a traffic network may lead to severe consequences, e.g., breakdowns in parts of the traffic network due to the propagation of congestion. Thus, means to analyze collectivity as proposed here are called for. Collective behavior is specific for certain time periods and regions, and seldom affects the system as a whole. Nevertheless, constructions in sensitive parts of the system, e.g., bridges, quickly bring larger parts of the motorway network close to collapses, drastically illustrating its vulnerability. The methods developed here are a first step to provide quantitative methods for assessing the collective effects leading to this type of vulnerability.

\section*{Acknowledgements}
\addcontentsline{toc}{section}{Acknowledgements} 

We gratefully acknowledge funding via the grant ``Korrelationen und deren Dynamik in Autobahnnetzen'', Deutsche Forschungsgemeinschaft (DFG, 418382724). We thank Strassen.NRW for providing the empirical traffic data.

\section*{Author contributions}

T.G. and M.S. proposed the research. S.W. and T.G. developed the methods of analysis. S.W. performed all the calculations. S.G. prepared the traffic data. S.W. and T.G. wrote the manuscript with input from M.S. and S.G.. All authors contributed equally to analyzing the results and reviewing the paper.

%\bibliographystyle{unsrt}
%\bibliography{reference.bib}

\begin{thebibliography}{10}
\addcontentsline{toc}{section}{References} 

\bibitem{Ladyman2013}
James Ladyman, James Lambert, and Karoline Wiesner.
\newblock What is a complex system?
\newblock {\em Eur. J. Philos. Sci.}, 3(1):33--67, 2013.

\bibitem{Ziemelis2001}
Karl Ziemelis.
\newblock Complex systems.
\newblock {\em Nature}, 410(6825):241--241, 2001.

\bibitem{Schmitt2013}
Thilo~A Schmitt, Desislava Chetalova, Rudi Sch{\"a}fer, and Thomas Guhr.
\newblock Non-stationarity in financial time series: Generic features and tail
  behavior.
\newblock {\em EPL (Europhysics Letters)}, 103(5):58003, 2013.

\bibitem{Stepanov2015}
Yuriy Stepanov, Philip Rinn, Thomas Guhr, Joachim Peinke, and Rudi Sch{\"a}fer.
\newblock Stability and hierarchy of quasi-stationary states: financial markets
  as an example.
\newblock {\em J. Stat. Mech: Theory Exp.}, 2015(8):P08011, 2015.

\bibitem{Wang2016}
Hao Wang, Teng Wu, Tianyou Tao, Aiqun Li, and Ahsan Kareem.
\newblock Measurements and analysis of non-stationary wind characteristics at
  sutong bridge in typhoon damrey.
\newblock {\em J. Wind. Eng. Ind. Aerodyn.}, 151:100--106, 2016.

\bibitem{Guhr1998}
Thomas Guhr, Axel M{\"u}ller-Groeling, and Hans~A Weidenm{\"u}ller.
\newblock Random-matrix theories in quantum physics: common concepts.
\newblock {\em Phys. Rep.}, 299(4-6):189--425, 1998.

\bibitem{Plerou2002}
Vasiliki Plerou, Parameswaran Gopikrishnan, Bernd Rosenow, Luis A~Nunes Amaral,
  Thomas Guhr, and H~Eugene Stanley.
\newblock Random matrix approach to cross correlations in financial data.
\newblock {\em Phys. Rev. E}, 65(6):066126, 2002.

\bibitem{Potestio2009}
Raffaello Potestio, Fabio Caccioli, and Pierpaolo Vivo.
\newblock Random matrix approach to collective behavior and bulk universality
  in protein dynamics.
\newblock {\em Phys. Rev. Lett.}, 103(26):268101, 2009.

\bibitem{Laloux1999}
Laurent Laloux, Pierre Cizeau, Jean-Philippe Bouchaud, and Marc Potters.
\newblock Noise dressing of financial correlation matrices.
\newblock {\em Phys. Rev. Lett.}, 83(7):1467, 1999.

\bibitem{Guhr2003}
Thomas Guhr and Bernd K\"alber.
\newblock A new method to estimate the noise in financial correlation matrices.
\newblock {\em J. Phys. A: Math. Gen.}, 36(12):3009, 2003.

\bibitem{Laloux2000}
Laurent Laloux, Pierre Cizeau, Marc Potters, and Jean-Philippe Bouchaud.
\newblock Random matrix theory and financial correlations.
\newblock {\em Int. J. Theor. Appl. Finance}, 3(03):391--397, 2000.

\bibitem{Gopikrishnan2001}
Parameswaran Gopikrishnan, Bernd Rosenow, Vasiliki Plerou, and H~Eugene
  Stanley.
\newblock Quantifying and interpreting collective behavior in financial
  markets.
\newblock {\em Phys. Rev. E}, 64(3):035106, 2001.

\bibitem{Wang2016a}
Shanshan Wang, Rudi Sch{\"a}fer, and Thomas Guhr.
\newblock Cross-response in correlated financial markets: individual stocks.
\newblock {\em Eur. Phys. J. B}, 89(4):105, 2016.

\bibitem{Wang2016b}
Shanshan Wang, Rudi Sch{\"a}fer, and Thomas Guhr.
\newblock Average cross-responses in correlated financial markets.
\newblock {\em Eur. Phys. J. B}, 89(9):207, 2016.

\bibitem{Benzaquen2017}
Michael Benzaquen, Iacopo Mastromatteo, Zoltan Eisler, and Jean-Philippe
  Bouchaud.
\newblock Dissecting cross-impact on stock markets: An empirical analysis.
\newblock {\em J. Stat. Mech: Theory Exp.}, 2017(2):023406, 2017.

\bibitem{Kerner2004}
Boris~S Kerner.
\newblock {\em The physics of traffic: empirical freeway pattern features,
  engineering applications, and theory}.
\newblock Springer, 2004.

\bibitem{Wang2018}
Shanshan Wang, Sebastian Neus{\"u}{\ss}, and Thomas Guhr.
\newblock Statistical properties of market collective responses.
\newblock {\em Eur. Phys. J. B}, 91(8):1--11, 2018.

\bibitem{Heckens2020}
Anton~J Heckens, Sebastian~M Krause, and Thomas Guhr.
\newblock Uncovering the dynamics of correlation structures relative to the
  collective market motion.
\newblock {\em J. Stat. Mech: Theory Exp.}, 2020(10):103402, 2020.

\bibitem{Nagel1992}
Kai Nagel and Michael Schreckenberg.
\newblock A cellular automaton model for freeway traffic.
\newblock {\em J. Phys. I}, 2(12):2221--2229, 1992.

\bibitem{Schadschneider1993}
Andreas Schadschneider and Michael Schreckenberg.
\newblock Cellular automation models and traffic flow.
\newblock {\em J. Phys. A: Math. Gen.}, 26(15):L679, 1993.

\bibitem{Lovaas1994}
Gunnar~G L{\o}v{\aa}s.
\newblock Modeling and simulation of pedestrian traffic flow.
\newblock {\em Transport. Res. B-Meth.}, 28(6):429--443, 1994.

\bibitem{Schreckenberg1995}
Michael Schreckenberg, Andreas Schadschneider, Kai Nagel, and Nobuyasu Ito.
\newblock Discrete stochastic models for traffic flow.
\newblock {\em Phys. Rev. E}, 51(4):2939, 1995.

\bibitem{Hoogendoorn2001}
Serge~P Hoogendoorn and Piet~HL Bovy.
\newblock State-of-the-art of vehicular traffic flow modelling.
\newblock {\em Proc. Inst. Mech. Eng., Part I: J. Syst. Control Eng.},
  215(4):283--303, 2001.

\bibitem{Wong2002}
GCK Wong and SC~Wong.
\newblock {A multi-class traffic flow model--an extension of LWR model with
  heterogeneous drivers}.
\newblock {\em Transp. Res. Part A Policy Pract.}, 36(9):827--841, 2002.

\bibitem{Fellendorf2010}
Martin Fellendorf and Peter Vortisch.
\newblock {Microscopic traffic flow simulator VISSIM}.
\newblock In {\em {Fundamentals of Traffic Simulation}}, pages 63--93.
  Springer, 2010.

\bibitem{Treiber2013}
Martin Treiber and Arne Kesting.
\newblock {\em {Traffic Flow Dynamics: Data, Models and Simulation}}.
\newblock Springer, 2013.

\bibitem{Kerner2002}
Boris~S Kerner.
\newblock Empirical macroscopic features of spatial-temporal traffic patterns
  at highway bottlenecks.
\newblock {\em Phys. Rev. E}, 65(4):046138, 2002.

\bibitem{Bertini2005}
Robert~L Bertini and Monica~T Leal.
\newblock Empirical study of traffic features at a freeway lane drop.
\newblock {\em J. Transp. Eng.}, 131(6):397--407, 2005.

\bibitem{Schonhof2007}
Martin Sch{\"o}nhof and Dirk Helbing.
\newblock Empirical features of congested traffic states and their implications
  for traffic modeling.
\newblock {\em Transp. Sci.}, 41(2):135--166, 2007.

\bibitem{Wang2020}
Shanshan Wang, Sebastian Gartzke, Michael Schreckenberg, and Thomas Guhr.
\newblock Quasi-stationary states in temporal correlations for traffic systems:
  Cologne orbital motorway as an example.
\newblock {\em J. Stat. Mech: Theory Exp.}, 2020(10):103404, 2020.

\bibitem{border}
{Bundesamt f\"ur Kartographie und Geod\"asie}.
\newblock {Verwaltungsgebiete 1:2 500 000, Stand 01.01. (VG2500)}.
\newblock
  \url{https://gdz.bkg.bund.de/index.php/default/verwaltungsgebiete-1-2-500-000-stand-01-01-vg2500.html},
  2020.

\bibitem{licence}
{Das Datenportal f\"ur Deutschland}.
\newblock {Data licence Germany -- attribution -- version 2.0}.
\newblock \url{http://www.govdata.de/dl-de/by-2-0}, 2021.

\bibitem{population}
{Statistische \"Amter des Bundes und der L\"ander, Deutschland}.
\newblock {Regionalatlas Deutschland}.
\newblock \url{https://regionalatlas.statistikportal.de}, 2021.

\bibitem{osmcopyright}
{OpenStreetMap}.
\newblock {Copyright and License}.
\newblock \url{https://www.openstreetmap.org/copyright}, 2021.

\bibitem{osm}
{Open Knowledge Foundation}.
\newblock {Open Data Commons Open Database License (ODbL) v1.0}.
\newblock \url{https://opendatacommons.org/licenses/odbl/1-0/}, 2021.

\bibitem{qgis}
{QGIS}.
\newblock {Documentation for QGIS 3.4}.
\newblock \url{https://docs.qgis.org/3.4/en/docs/}, 2020.

\bibitem{Marchenko1967}
Vladimir~Alexandrovich Marchenko and Leonid~Andreevich Pastur.
\newblock Distribution of eigenvalues for some sets of random matrices.
\newblock {\em Mat. Sb.}, 114(4):507--536, 1967.

\bibitem{Song2011}
Dong-Ming Song, Michele Tumminello, Wei-Xing Zhou, and Rosario~N Mantegna.
\newblock Evolution of worldwide stock markets, correlation structure, and
  correlation-based graphs.
\newblock {\em Phys. Rev. E}, 84(2):026108, 2011.

\bibitem{Pharasi2019}
Hirdesh~K Pharasi, Kiran Sharma, Anirban Chakraborti, and Thomas~H Seligman.
\newblock Complex market dynamics in the light of random matrix theory.
\newblock In {\em New Perspectives and Challenges in Econophysics and
  Sociophysics}, pages 13--34. Springer, 2019.

\bibitem{Anderson1952}
Theodore~W Anderson and Donald~A Darling.
\newblock Asymptotic theory of certain "goodness of fit" criteria based on
  stochastic processes.
\newblock {\em Ann. Math. Stat.}, pages 193--212, 1952.

\end{thebibliography}

\end{document}